\documentclass[fleqn,usenatbib]{mnras}

\usepackage{newtxtext,newtxmath}

\usepackage[T1]{fontenc}

\DeclareRobustCommand{\VAN}[3]{#2}
\let\VANthebibliography\thebibliography
\def\thebibliography{\DeclareRobustCommand{\VAN}[3]{##3}\VANthebibliography}

\usepackage{graphicx}	
\usepackage{amsmath}	

\title[Mapping Outflows in Mrk~1486]{DUVET Survey: Mapping Outflows in the Metal-Poor Starburst Mrk~1486}

\author[D. K. McPherson et al.]{
Daniel K. McPherson,$^{1, 2}$\thanks{E-mail: dmcpherson@swin.edu.au}
Deanne B. Fisher,$^{1, 2}$
Nikole M. Nielsen,$^{1,2}$
Glenn G. Kacprzak,$^{1,2}$\newauthor
Bronwyn Reichardt Chu,$^{1,2}$
Alex J. Cameron,$^{3}$
Alberto D. Bolatto,$^{4}$
John Chisholm,$^{5}$\newauthor
Drummond B. Fielding,$^{6}$
Danielle Berg,$^{5}$
Rodrigo Herrera-Camus,$^{7}$
Miao Li,$^{8}$\newauthor
Ryan J. Rickards Vaught,$^{9}$
and Karin Sandstrom$^{9}$
\\
$^{1}$Centre for Astrophysics and Supercomputing, Swinburne University of Technology, Hawthorn, Victoria 3122, Australia\\
$^{2}$ARC Centre of Excellence for All Sky Astrophysics in 3 Dimensions (ASTRO 3D), Australia\\
$^{3}$Sub-department of Astrophysics, University of Oxford, Keble Road, Oxford OX1 3RH, United Kingdom\\
$^{4}$Department of Astronomy, University of Maryland, College Park, MD 20742, USA\\
$^{5}$Department of Astronomy, University of Texas, Austin, TX 78712, USA\\
$^{6}$Center for Computational Astrophysics, Flatiron Institute, 162 Fifth Avenue, New York, NY 10010, USA\\
$^{7}$Departamento de Astronom\'ia, Universidad de Concepci\'on, Barrio Universitario, Concepci\'on, Chile\\
$^{8}$Department of Physics, Zhejiang University, 866 Yuhangtang Road, Hangzhou, 310058, China\\
$^{9}$Center for Astrophysics and Space Sciences, Department of Physics, University of California, San Diego, CA, USA\\
}

\date{Accepted XXX. Received YYY; in original form ZZZ}

\pubyear{2023}

\begin{document}
\label{firstpage}
\pagerange{\pageref{firstpage}--\pageref{lastpage}}
\maketitle

\begin{abstract}
We present a method to characterize star-formation driven outflows from edge-on galaxies and apply this method to the metal-poor starburst galaxy, Mrk~1486. Our method uses the distribution of emission line flux (from H$\beta$ and [OIII]~5007) to identify the location of the outflow and measure the extent above the disk, the opening angle, and the transverse kinematics. We show that this simple technique recovers a similar distribution of the outflow without requiring complex modelling of line-splitting or multi-Gaussian components, and is therefore applicable to lower spectral resolution data. In Mrk~1486 we observe an asymmetric outflow in both the location of the peak flux and total flux from each lobe. We estimate an opening angle of $17-37^{\circ}$ depending on the method and assumptions adopted. Within the minor axis outflows, we estimate a total mass outflow rate of $\sim$2.5~M$_{\odot}$~yr$^{-1}$, which corresponds to a mass loading factor of $\eta=0.7$. We observe a non-negligible amount of flux from ionized gas outflowing along the edge of the disk (perpendicular to the biconical components), with a mass outflow rate $\sim0.9$~M$_{\odot}$~yr$^{-1}$. Our results are intended to demonstrate a method that can be applied to high-throughput, low spectral resolution observations, such as narrow band filters or low spectral resolution IFS that may be more able to recover the faint emission from outflows. 
\end{abstract}

\begin{keywords}
galaxies: Mrk~1486 -- galaxies: evolution -- galaxies: starburst -- galaxies: star formation
\end{keywords}



\section{Introduction} \label{sec:intro}

Galaxy-scale outflows are ubiquitous in high star-formation rate galaxies in both the local Universe \citep[e.g.][]{Veilleux2005} and at higher redshift \citep{Rubin2014,Steidel2010}. There is a consensus view that star formation driven outflows are a necessary component for models of galaxy evolution to reproduce observations of galaxy properties such as the stellar mass function \citep{Somerville2015,Naab2017,Pillepich2018,Forster2019}. Current models of galaxy evolution propose that these outflows regulate star-formation by removing star-forming material from the galaxy disk \citep{Oppenheimer2008} and enriching the surrounding circumgalactic medium (CGM) with higher metallicity gas \citep[e.g.][]{Peroux2020}. This enrichment of outflows has, for the first time, been directly mapped by \citet{Cameron2021}. The direct mapping of outflows is now allowing us to make more accurate measurements of mass outflow rates.

The observation and characterization of star-formation driven outflows has historically been limited by the extremely low surface brightness of extraplanar gas \citep{Tumlinson2017}. By far the most extensively studied of these is M82 \citep{Lopez2020, Leroy2015, Shopbell1998, Westmoquette2009}. Outflows have also been directly imaged in NGC~1482 \citep{Veilleux2002}, NGC~253 \citep{Bolatto2013}, and in a sample of nearby outflow candidates \citep{Veilleux2003}. \citet{Concas2022} finds evidence for outflows in massive galaxies at $z\sim2$ from the KLEVER survey, but no indication of such outflows in lower-mass galaxies, which may be due to the lower S/N on fainter targets. A study on 19 dwarf galaxies from the Dwalin sample found evidence of outflows but with very low mass outflow rates \citep{Marasco2022}. This however represents a small sample of galaxies. Additionally, there are a number of methods used in studying outflows, and the definition of mass outflow rate changes between studies. This makes it difficult to compare results between samples and to compare results to simulations. 
    
Models of galaxy evolution make direct predictions of the mass outflow rate, $\dot{M}_{\rm out}$, of galaxies based on basic properties such as stellar mass, SFR and gas fraction \citep[e.g.][]{Nelson2019, Hayward2017}. The mass outflow rate is therefore a critical parameter for observations to recover. The mass outflow rate is defined as: 
\begin{equation}
    \label{eq:mout}
    \dot{M}_{\rm out} = \Omega C_f \mu m_p N_{H} R_{\rm out} v_{\rm out}.
\end{equation}
In the above equation, $\Omega$ is the opening angle of the outflowing gas. $C_f$ is the covering fraction of the outflow. $\mu$ is the mass per H nucleus, accounting for the relative He abundance. $m_{p}$ is the proton mass. $N_{H}$ is the column density of outflowing gas. R$_{\rm out}$ is the radial extent of the outflow, and the velocity of the outflow is represented by $v_{\rm out}$. There is difficulty in measuring all these parameters, and they are frequently assumed. This introduces potentially large systematic uncertainties into determinations of the mass outflow rate.

The most common method to derive outflow properties is the technique of decomposing spectral lines into an outflow and a systemic line, which can be performed on large samples of galaxies with either absorption or emission lines \citep[e.g.][]{Rubin2014, Chisholm2015, Heckman2015, Forster2019, ReichardtChu2022}. In these works, the geometric parameters (opening angle, covering fraction, and outflow radius) must be assumed, rather than directly measured. In the nearby starburst M82, the outflow has been measured to extend $\sim10-15$~kpc with a base width of $\sim0.5$~kpc and opening angle $25^{\circ}$ \citep{Shopbell1998}. We can measure these properties in M82 due to its proximity, which allows us to accurately measure the physical sizes of the faint emission. Many studies assume covering fractions of $\sim$0.8-1 \citep{Chisholm2015, Heckman2015}, but there is a large variation between galaxies \citep{Martin2005}, which may depend highly on galaxy morphology. 

Assumptions on outflow radius range from 100~pc to kiloparsecs \citep[e.g.][]{Chisholm2015, Forster2019}. As these outflow properties are direct inputs for the mass outflow rate, an incomplete understanding of how they scale with galaxy properties and morphologies results in poorly constrained mass outflow rates. We are especially lacking in understanding how properties like the covering fraction or opening angle may vary with galaxy mass and SFR.

Image-slicer integral field units (IFUs) such as VLT/MUSE and Keck/KCWI make direct imaging of the morphologies and extents of extraplanar gas possible in moderate-sized samples of galaxies. Using VLT/MUSE, to this end effort has been made to study individual outflows in more extreme and distant galaxies \citep{Rupke2019, Burchett2021, Shaban2022, Zabl2021}. A result of these efforts has been the determination of the large extent of the outflows in these intermediate ($z\sim0.5$) to high redshift systems ($z\sim1.7$) with the detection of emission in MgII extending to $\sim30$~kpc \citep{Burchett2021, Shaban2022, Zabl2021} and [OII] to $\sim40$~kpc. A limitation in many of these higher redshift studies (that don't make use of gravitational lensing as in \citet{Shaban2022}) is the reduced spatial resolution. This makes difficult the determination of wind properties such as opening angle and outflow radius close to the galaxy disk. Additionally, image-slicer IFUs introduce the new challenge of separating outflowing gas from the surrounding halo gas. The ability to measure the parameters $\Omega,~C_{f},$ and $R_{\rm out}$ in Eq.~\ref{eq:mout} in more comprehensive samples of galaxies is central to determining how mass outflow rates vary with galaxy properties. In order to do this we must first have methods to determine these properties that can be systematically applied to deep IFU datasets. 

The DUVET (Deep near-UV observations of Entrained gas in Turbulent galaxies) survey is an IFU survey on KCWI with sub-kpc spatial resolution (in contrast to previous IFU surveys) observations of 27 starbursting, low-redshift galaxies ($z\sim0.03$) \citep[e.g.][]{Cameron2021,ReichardtChu2022}. Galaxies selected for the sample have star-formation rates at least 5 times the main-sequence value for their stellar mass. In addition the survey requires that galaxies have morphologies and kinematics consistent with a disk. Amongst the DUVET galaxies are several edge-on systems, with extended minor axis emission. The subject of this paper is a detailed analysis of one edge-on outflow galaxy from the DUVET survey, Mrk~1486 as a case-study for characterising the outflow emission.

Throughout this paper we assume a flat $\Lambda$CDM cosmology with
$H_{0}=69.3$~km~Mpc$^{-1}$~s$^{-1}$, $\Omega_{m}=0.3$, and $\Omega_{\Lambda}=0.7$. All wavelengths quoted are rest-frame wavelengths. 

\begin{figure}
    \centering
    \includegraphics[width=\linewidth]{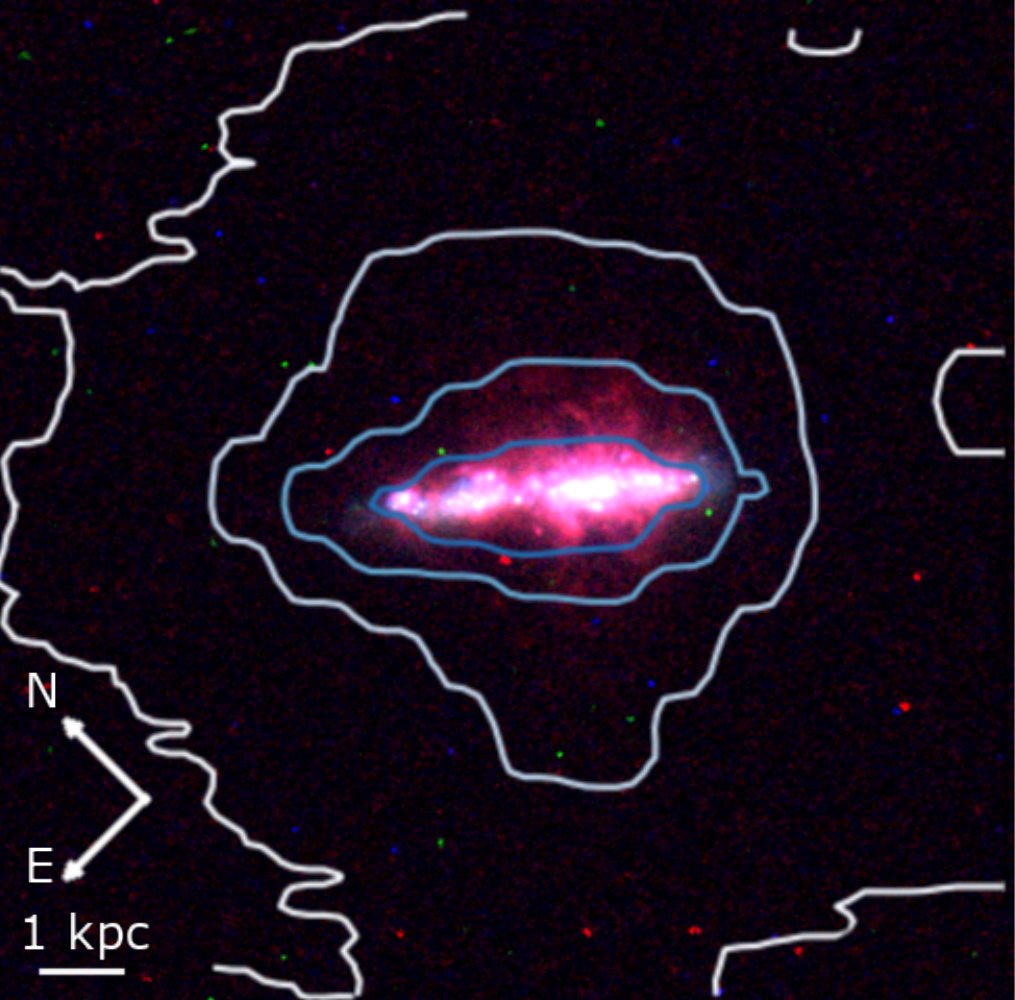}
    \caption{$19''\times19''$ HST image of Mrk~1486 combining F336W (blue), F438W (green), and F673N (red). The young stars establish an edge-on disk. The H$\alpha$ emission (red) shows filamentary structures extending above and below the plane of the disk, indicating the outflow described in \citet{Duval2016} and recently in \citet{Cameron2021}. Overlaid are contours in decadal steps showing [OIII]~$\lambda5007$ emission detected with KCWI. Note that the brightness and contrast is set to make the outflow visible.}
    \label{fig:three_colour}
\end{figure}

\section{Observations and Data Reduction} \label{sec:satfix}

\subsection{Target: Mrk 1486}
Mrk~1486 is an edge-on disk, with inclination $85^{\circ}$ \citep{Chisholm2015}. It has a redshift of $z=0.03383$ and stellar mass $\log(M_{\star}/M_{\odot})=9.3\pm0.2$. It is a 5x outlier above the star-formation rate main-sequence, similar to other galaxies with observed strong star-formation driven outflows, with star formation rate ${\rm SFR}=3.6\pm0.7$~M$_{\odot}$~yr$^{-1}$ \citep{Chisholm2018b}. Mrk~1486 has a low ISM metallicity of $12 + \log(\rm{O/H})=7.8$ \citep{Ostlin2014}. Mrk~1486 also hosts a bipolar outflow \citep{Duval2016}, visible as extended filamentary structures in HST {H$\alpha$} imaging (fig.~\ref{fig:three_colour}). These properties make Mrk~1486 an ideal target to study basic outflow properties. 

\subsection{KCWI Observations}
Observations of Mrk~1486 were taken on 2020 March 22 UT under sub-arcsecond seeing conditions ($\sim0\farcs7$ at 5000~\AA) with Keck/KCWI \citep{Morrissey2018} using the large IFU slicer setting giving a spatial sampling of $0\farcs29\times1\farcs35$ and a $20''\times33''$ field of view. Two configurations on the BM dispersion grating were used, a ``blue" configuration with a central wavelength of 4180~{\AA}, and a ``red" configuration with a central wavelength of 4850~{\AA}. This allowed for continuous spectral coverage from 3731~{\AA} -- 5284~{\AA} with spectral resolution $R\sim2000$.

Our project aims require tracking bright emission lines from the galaxy center to the faint gas in the outflow and surrounding region. The KCWI detector rapidly saturates with bright [OIII]~$\lambda\lambda4959,5007$ doublet and H$\beta$ from the starburst. The $\sim$1~minute readout time makes a large number of short exposures time-prohibitive to reach our aims. We, therefore, used a combination of long and short exposure times to avoid saturating the emission lines in the bright galaxy center while measuring faint emission in the galaxy outskirts. In the red configuration, nine exposures were taken, seven long ($6\times300$~s and $1\times400$~s) and two short ($2\times30$~s). In the blue configuration seven 300~s exposures were taken. A half-slice dither was used in both configurations to increase the spatial sampling. To adequately remove the sky, we obtained separate sky fields in the two configurations, where a $600$~s exposure was obtained in the red configuration directly before the science exposures and a $300$~s exposure was obtained in the blue configuration directly after the science exposures.

\subsection{Data Reduction}
The data were reduced with the IDL version of the KCWI Data Reduction Pipeline v1.1.0\footnote{\url{https://github.com/Keck-DataReductionPipelines/KcwiDRP}} using the standard settings with the separate sky fields noted above. The standard star Feige92 was used to flux calibrate the exposures in the final processing step. 

Before combining images, we align each datacube together using the H$\gamma$ emission line, which is unsaturated in all spaxels and covered in both the red and blue spectral settings. This accounts for small scale imperfections in the WCS. The alignment is carried out using an iterative minimisation method for each line-map, in which the reference position of the fields are adjusted and the H$\gamma$ flux is compared in each pixel. The position that results in the minimum average residual across the galaxy is chosen.       

Our chosen combination of exposure times results in long exposures with a few bright lines that saturated in the center, and short exposures that are not sufficiently deep to probe outflow gas nor can they probe fainter spectral features \citep[e.g.][]{Cameron2021}. We, therefore, developed a method to combine these two data sets with a preference toward using the longer exposures when there is no evidence for saturation within a reasonable bandpass. The [OIII]~$\lambda5007/\lambda4959$ ratio has a fixed value of $3$ \citep{Osterbrock2006}, and is close enough in wavelength to not be significantly impacted by extinction. We can use this as an indicator of saturation in each spaxel spectrum. In the pre-flux calibrated cubes (from step \texttt{kcwi\_stage7dar}) we use the $\lambda5007/\lambda4959$ ratio to determine the counts at which saturation is occurring. We found this to be at $\sim$5500 counts.

Saturation was not detected in any exposures in the blue configuration, and thus no short exposures were taken, and only long exposures were used. These exposures were reprojected to produce $0\farcs29\times0\farcs29$ square spaxels with the python package \textsc{Montage}\footnote{\url{http://montage.ipac.caltech.edu/}}. This size was chosen based on the length of the smaller edge of the original rectangular spaxels. The reprojected images were then co-added with \textsc{Montage}. During reprojection we set \texttt{drizzle=1.0}, \texttt{energyMode=True}, and scaled the flux with the \texttt{fluxScale} parameter set to the ratio between the rectangular and square spaxel sizes. Variance cubes for the blue configuration were also reprojected and co-added in the same manner.

For the red configuration, there was no saturation detected in the short exposures. These were co-added using \textsc{Montage}, and variance cubes were scaled as above. For the long exposures, saturation was detected near the galaxy center in the H$\beta$, [OIII]~$\lambda4959$, and [OIII]~$\lambda5007$ emission lines. Where a saturated emission line was detected in a spaxel in a long exposure, a 20~{\AA} wavelength region centered on the emission line was replaced with the corresponding wavelength region in the combined short exposure in both the flux cube and the variance cube. Light was found to bleed from saturated spaxels to spatially adjacent ones, resulting in deviations of the [OIII]~$\lambda5007/\lambda4959$ ratio from the theoretically predicted value well below the determined count threshold to detect saturation. To correct for this effect this replacement was also carried out in a one spaxel annulus surrounding the spaxel where saturation was detected. The individual long exposures were then reprojected to produce square spaxels and co-added, in the same manner as the blue exposures. 

The final images have a total integration time of 2100~s in the blue setting across the entire field, 2200~s in the red setting off the galaxy disk where the saturation correction was not applied, and 60s on the disk where the saturation correction was applied. The images cover the disk of Mrk~1486, and extend to a minor axis distance of $6.9$~kpc in both the NW and SE, and a major axis distance of $12.2$~kpc in the NE and $5.5$~kpc in the SW, with a $3\sigma$ surface brightness limit of $2.3\times10^{-18}$~erg~s$^{-1}$~cm$^{-2}$~{\AA}$^{-1}$~arcsec$^{-1}$.

\section{Data Analysis} \label{sec:analysis}

\subsection{Continuum Subtraction} \label{subsec:contsub}

Our observations cover both the galaxy disk and the surrounding extraplanar regions. As a result, our field of view includes some regions in which we expect continuum emission (near the galaxy) and regions further from the galaxy with little-to-no continuum emission. Continuum subtraction thus cannot be applied across the whole cube. For those spaxels in the galaxy we use standard methods of continuum removal with pPXF \citep{Cappellari2017} with BPASS templates \citep{Stanway2018}. To determine those spaxels with sufficient continuum flux to employ pPXF, we estimate the continuum signal-to-noise in each spaxel. Continuum signal-to-noise is estimated by summing the flux in the band, after masking emission lines. In spaxels where we detected continuum flux at a signal-to-noise level of 3 or greater the continuum was fitted with pPXF. We note that in all spaxels, independant of continuum signal-to-noise, our Gaussian fitting includes a constant offset determined near to the emission line, which corrects for local imperfections in continuum fitting.  

The templates used included binary systems, and were based on a broken powerlaw initial mass function, with a slope of $-1.3$ between 0.1~M$_{\odot}$ and 1~M$_{\odot}$, a slope of $-2.35$ above 1~M$_{\odot}$ and an upper mass limit of 300~M$_{\odot}$. The reddening of the galaxy spectra caused by Milky Way foreground extinction was corrected for using the \citet{Calzetti2001} extinction curve. The internal extinction in the galaxy was then determined using the {H$\beta$/H$\gamma$} ratio and corrected also using a \citet{Calzetti2001} extinction law. 

\subsection{Emission Line Fitting} \label{subsec:linefit}

We carry out emission line fitting with two separate settings, a single Gaussian fit to all spaxels and a second run in which multiple Gaussians are fit to each spaxel. Our  emission line fitting software is built on work done by \citet{ReichardtChu2022} and uses the python package \textsc{threadcount}\footnote{\url{https://github.com/astrodee/threadcount}}. We do not make a correction for the instrumental dispersion on the line width, as it would be mostly negligible for any dispersions <100~km~s$^{-1}$. 

We adopt a Bayesian method to decide spaxel-by-spaxel how many Gaussian components are needed. The software fits each spaxel with separate models of one, two, and three Gaussian components for the [OIII]~$\lambda5007$ emission line. We use the Python package \texttt{lmfit} \citep{lmfit} with the \texttt{nelder} minimization algorithm. The \texttt{leastsquares} minimization algorithm was tested, but was less reliable at fitting complex emission lines in the lower surface brightness regions due to the large number of local minima. We use the built in Akaike Information Criterion (AIC) function in \textsc{threadcount} to determine the validity of a model. A model is determined to be more significant for an AIC difference of 150. \citet{ReichardtChu2022} discusses that the typically adopted difference of 10 results in all spaxels having multiple components. This is likely due to the reality that galaxy emission lines are not perfect Gaussians, and our very high S/N data identifies these small deviations. We chose the value of  150 based on visual inspection of characteristic spaxels. A quantified analysis of appropriate AIC values is in progress (Reichardt Chu et al.\ {\em in prep}). In our calculation, a 2-Gaussian fit must have an AIC that is 150 smaller than the AIC of the 1-Gaussian fit. Similarly a 3-Gaussian model must have an AIC value that is 150 smaller than both the 2-Gaussian and 1-Gaussian models. The user is prompted to judge individual fits when parameters fall into a range that is deemed uncertain. For example, if the central wavelengths of two  differ by less than 0.5~{\AA} (the spectral resolution of the measurements) and if the ratio of the fluxes is lower than 0.25 the user double checks the decision made by the software. We then store both a single Gaussian fit for every spaxel and separately a multi-Gaussian model for each spaxel as decided by the AIC system.

\section{Decomposing Outflows from surrounding gas in Edge-On Galaxies}\label{sec:decompose}

In the following sections we present \textsc{threadcount} and our method for distinguishing the region of extraplanar gas that is likely outflow, where we determine the outflow region based on the observed surface brightness alone. As such our method can be more easily applied to large data sets of either deep IFU observations of emission lines or narrow band photometric observations of outflowing galaxies. We demonstrate that the determination of the outflow region using our method corresponds well with the outflow region that would be determined from traditional kinematic arguments.

\subsection{Identifying Outflows with Emission Line Surface Brightness} \label{subsec:outregionmethod}

\begin{figure*}
    \centering
    \includegraphics[width=\textwidth]{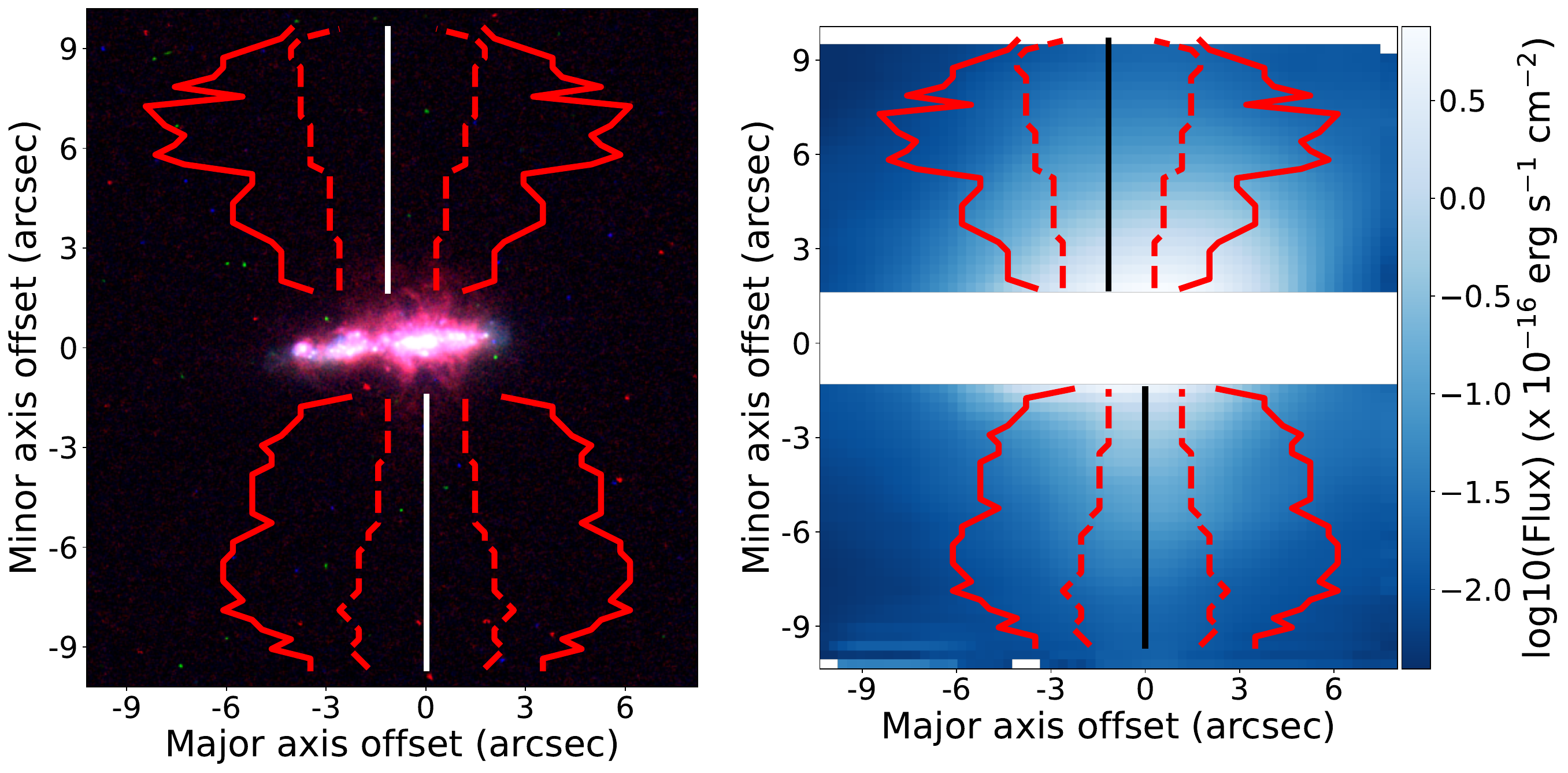}
    \caption{The outflow region (red lines) determined by \textsc{threadcount} overlaid on the HST image of this galaxy from Fig.~\ref{fig:three_colour} (\emph{left}) and [OIII]~$\lambda5007$ emission line flux (\emph{right}). The single Gaussian dispersion is used to define the disk region as distinct from the surrounding gas. Then the brightest spaxel in each row parallel to the galaxy disk is determined and the median position of these spaxels is taken as the central outflow axis (black vertical line). We then determine the 50\% and 90\% widths on either side of this line in terms of surface brightness (red dashed and solid lines, respectively) and define this as the outflow region.} 
    \label{fig:outflow_region}
\end{figure*}

\begin{figure}
    \centering
    \includegraphics[width=\linewidth]{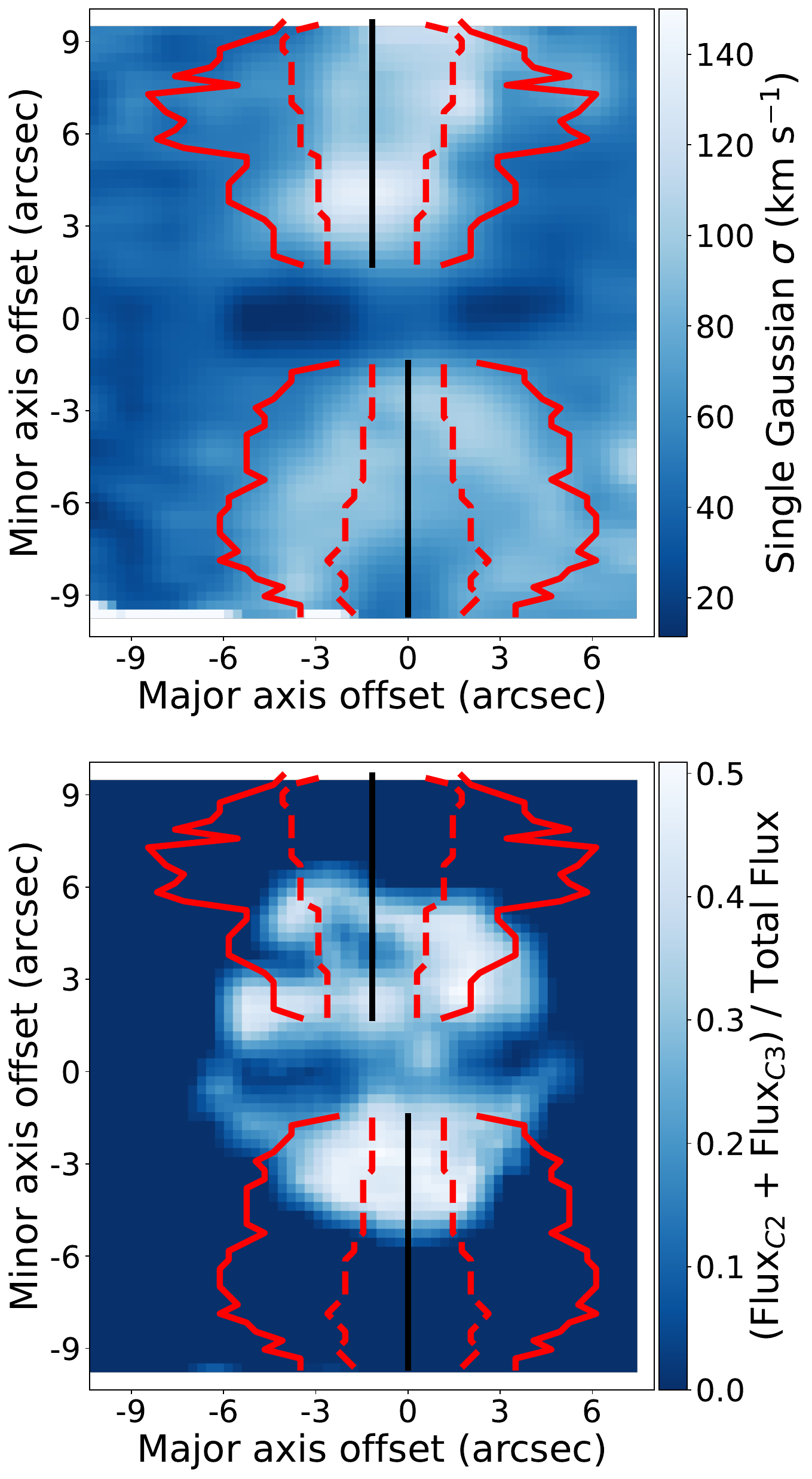}
    \caption{The \textsc{threadcount} determined outflow region overlaid on a $5\times5$ spaxel averaged map of the [OIII]~$\lambda5007$ single Gaussian dispersion (\emph{top}). The determined outflow region agrees strongly with the high-dispersion, minor axis regions. This outflow region is also overlaid on a $5\times5$ spaxel averaged map of the ratio of the flux in the two least bright components to the total flux in the multi-Gaussian component model for [OIII]~$\lambda5007$ (\emph{bottom}). High values for this ratio indicate regions where the high-dispersion, minor axis regions are not solely tracing an increase in turbulence, but are the result of an ordered, expanding shell of gas. Close to the disk where there is sufficient signal-to-noise to justify these higher-order fits the region we determine to be outflow from our surface brightness method contains the regions with high values for this flux ratio. Further from the disk, our fitting software prefers simpler fits as a result of the lower signal-to-noise. These maps indicate that the outflow region determined via our surface brightness based method agrees strongly with outflow determinations based on gas kinematics.} 
    \label{fig:line_splitting}
\end{figure}

Before determining the outflow regions and their properties we have to exclude the regions dominated by the disk of Mrk~1486. In our observations we find that along the minor axis the continuum drops to negligible values within a full width of order 1.5~kpc. We note that this is roughly equivalent to twice the seeing limit of our observations, and is probably what is setting this. Indeed, \citet{Duval2016} use the HST imaging in Mrk~1486 to define a disk scale-height of $\pm$0.25~kpc.  Moreover, we find that the linewidth, as traced by the single-Gaussian velocity dispersion shows a significant increase at the $z$ distance of $\pm$1~kpc from the disk midplane (as defined by the peak of the continuum brightness). We therefore make a cut of $\pm$1~kpc to separate the ``disk" from the ``outflow" of the galaxy. We note that the outflow likely originates from a star-forming region that is smaller \citep{Chisholm2018} and that our thickness is likely seeing-convolved. This distance is a rough approximation of where the emission line brightness becomes dominated by outflows and is similar to what was used in \citet{Leroy2015} for M~82. 

To define our outflow search area, the software must first identify a central axis for each side of the outflow. To do this, \textsc{threadcount} steps outward from the galaxy, finding the brightest spaxel within each strip parallel to the galaxy disk in terms of surface brightness for a chosen emission line. The [OIII]~$\lambda5007$ emission line is used throughout this work, as it is the brightest line in this galaxy, however any observed emission line can be chosen. The software stops iterating when no spaxels exist with brightness greater than the noise threshold given. In principle, this could be used to judge the outflow extent. This is, however, an observationally-limited method, and in the case of Mrk~1486, the emission extends to the edge of the KCWI field-of-view.  The median position of all brightest spaxels in each row is used to define a line perpendicular to the galaxy disk. This is then taken to be the center of the outflow. The center is allowed to vary on separate sides of the galaxy. The software then uses the center to determine the 50\% and 90\% widths surrounding this center line based on surface brightness, and we take this to be our outflow region. The red lines in Fig.~\ref{fig:outflow_region} show the outflow region determined by this process overlaid on the HST image and the [OIII]~$\lambda5007$ emission line flux map.

We note that this method determines the shape of the outflow purely by empirical photometric measurements alone, and is model-independent. The benefit of this method is that it can be applied to data from low spectral resolution instruments, such as the low-resolution PRISM mode of NIRSpec/IFU on JWST, or a narrow-band filter. It is also not reliant on assumptions about the underlying velocity structure of the emission. 

We find that at its widest point the 90\% surface brightness contour reaches a width of 9.9~kpc while the core of the outflow (the 50\% surface brightness contour) reaches a maximum width of 4~kpc. The diameter that contains 90\% of the $i$-band flux of the starlight in the disk on the major axis is 4.1~kpc. This implies that the 50\% width of the outflow reaches a comparable width as the starlight in the disk. \citet{Duval2016} shows that extraplanar Ly~$\alpha$ extends across more of the disk in HST images, which may be consistent with our observations. The outflow in Mrk~1486 is significantly wider than the outflow in M~82, in which \citet{Shopbell1998} estimates a diameter of 0.4-0.6~kpc. This may be due to a difference in the launching region of the outflow of Mrk~1486. The outflow of M~82 is known to be centrally-concentrated and contained to a region of order $\sim$1~kpc in the center of the galaxy. This is similar to that of NGC~253, which likewise has a central starburst surrounded by a lower SFR disk. In Mrk~1486 we observe $\Sigma_{\rm SFR}>0.1$~M$_{\odot}$~yr$^{-1}$~kpc$^{-2}$ across the entire disk, which suggests that there is sufficient star formation to drive a wind \citep{Heckman2015}.

\subsection{Comparing our method to kinematic tracers of outflows}
\label{subsec:kinematicoutflow}

Figure \ref{fig:line_splitting} compares the outflow region determined via this surface brightness based method with kinematic tracers of the outflow. The top panel of Fig.~\ref{fig:line_splitting} shows the outflow region overlaid on a map of the single-Gaussian velocity dispersion. \citet{Ho2016} argue that the velocity dispersion of gas is a strong indicator of outflows. Similar conclusions can be drawn from \citet{Westmoquette2009}.  The outflow regions determined from our surface brightness based method contain the minor axis regions with higher velocity dispersion. An average velocity dispersion just inside the contour is $\sim$90-100~km~s$^{-1}$, and this drops to 40-60~km~s$^{-1}$ just outside the boundary, which is a decline of 30-40~km~s$^{-1}$. We interpret this as indicating that our surface brightness based determination of the outflow region is identifying a kinematically-distinct component of the gas. This is then consistent with a minor axis outflow. 

In the lower panel of fig.~\ref{fig:line_splitting} we compare our outflow determination with results from multi-component fitting. The value in each spaxel is an emission line ratio of the fainter two Gaussian components divided by the total flux. Within a spaxel the flux is described as $F_{\rm total} = F_{\rm G1}+ F_{\rm G2}+ F_{\rm G3}$, where $F_{\rm G1}$ is the flux of the brightest Gaussian, and so on for the others. The figure then plots $(F_{\rm G2}+ F_{\rm G3})/F_{\rm total}$, which is a metric of the presence of multiple Gaussian components. For spaxels where two components are fit, we simply treat $F_{\rm G3}=0$. Where only a single component is fit we have $F_{\rm G2}=F_{\rm G3}=0$. We note that a BIC incorporates signal-to-noise in the choice. Lower S/N spaxels are less likely to identify multiple components, which is the primary reason the multiple components are not strong beyond $\pm$6~arcsec. Inside the high S/N region we again find agreement between the strongest multiple component behavior with the outflow contours, overplotted in the figure. We note that so-called ``line splitting" is a key feature of many outflows \citep[e.g.][]{Westmoquette2009}.  Close to the disk we find that as much as half of the detected [OIII]~$\lambda5007$ flux is in these additional fit components. These regions also exhibit high values for the single Gaussian velocity dispersion. This suggests that the velocity dispersion in these regions is not solely tracing an increase in turbulence, but is consistent with expanding, semi-transparent shells of gas, which is the typical model of a conical outflow. The outflow region determined from surface brightness surrounds these regions with significant secondary and tertiary fit components. Our surface brightness based method for determining the outflow region therefore recovers the same region as kinematic metrics such as single-Gaussian dispersion and multi-component fitting. 

Overall, we find that identifying the outflow in Mrk~1486 with only surface brightness photometry recovers a similar region as more complex kinematic methods. 

\subsection{Multiple estimates of outflow opening angle} \label{subsec:idoutflow}

In this subsection we compare the opening angle measured using the geometry determined with the surface brightness method to that determined from the line-splitting. We note this is a heavily simplified interpretation of the opening angle, and it is intended to be both an empirically-based method that can be applied without need for significant modelling and tractable for comparison of samples of outflows. While \citet{Herenz2023}, for example, offers a method that relies on a toy-model applied to the outflow of SBS~0335-52E using two limb filaments, a simple geometric method applied to emission line surface brightness contours to determine the opening angle derives naturally from our outflow identification method. The opening angle in edge-on systems is typically determined using the velocity difference between the multiple components of an emission line, so-called ``line splitting" measurements. It is not known how common, or easily observed, line splitting is. It requires a minimum ability to resolve the components of the emission lines, and therefore is precluded from use in observations with low spectral resolution. Moreover, line splitting determination of the opening angle requires an assumption about the outflow velocity perpendicular to the disk, which is highly uncertain in edge-on systems.

To estimate the opening angle geometrically, we fit the 50\% contour of the outflow (the dashed line in figs.~\ref{fig:outflow_region} and \ref{fig:line_splitting}) with a straight line. We fit the NW and SE lobes separately. We then determine the angle between the fit line and the outflow central axis by measuring the arctangent of the slope of the fit. The opening angle is generally reported as the full width of the cone. Therefore, twice the angle between that fitted line and the outflow central axis is comparable to the opening angle of the outflow. Errors in opening angle are calculated by first running Monte Carlo simulations of major axis offsets for a given minor axis offset taking the spatial resolution of the image as the error in the major axis position. We then fit each simulation, and take the standard deviation in fit slope as the error in our measured slope. This error is then propagated to an error in the opening angle. We provide the full opening angle for the outflow shapes and calculation methods discussed below in Table~\ref{tab:outwidths}. With this method we measure an opening angle at the 50\% width of 19.81~$\pm$~0.01~$^{\circ}$ in the NW region and 17.7~$\pm$~0.02~$^{\circ}$ in the SE region assuming the frustrumal\footnote{A truncated cone.} geometry seem in figs.~\ref{fig:outflow_region} and \ref{fig:line_splitting}. At the 90\% width for these same angles, we measure 43.94~$\pm$~0.01~$^{\circ}$ in the NW region and 25.12~$\pm$~0.02~$^{\circ}$ in the SE region, though the dramatic changes in the major axis position of the 90\% contour make these angles less reliable as an indicator of the overall outflow opening angle.

We can compare this geometric method to the more common line-splitting determination of opening angle. We measure the mean line separation within $\pm1$~kpc of the central outflow axis, extending from 1 to 4~kpc minor axis offset from the galaxy. This excludes the galaxy disk and includes only regions of sufficient S/N to detect line splitting. In spaxels where the one component model was selected, the line separation is taken to be 0~km~s$^{-1}$. Where a two-component model is selected, the line separation is taken to be the difference between the Gaussian mean of each component. Where a three-component model is selected, the line separation is taken as the difference between the means of the most blueshifted and the most redshifted components. Assuming an expanding, biconical minor axis outflow we can use our measurements of line separation in these minor axis regions to constrain the opening angle of the cone ($\theta$) with
\begin{equation}
    \label{eq:openingangle}
    \theta=2\arcsin \left(\frac{\Delta v}{2v_{\rm out}}\right)
\end{equation}
from a direct geometric argument. Here $\Delta v$ is the observed separation between emission line components, and $v_{\rm out}$ is the outflow velocity. Based on the star-formation rate and stellar mass of the galaxy, there is precedent in the literature for outflow velocities in the range $v_{\rm out}\approx150-450$~km~s$^{-1}$ \citep{Chisholm2015, Heckman2015}. We assume the median value of $v_{\rm out}=300$~km~s$^{-1}$ in our calculation. With this line-splitting method we derive an opening angle of $\sim17^{\circ}$ in both the NW and SE regions. This is in strong agreement with the opening angle determined via our photometric method.

However, this line-splitting method for determining the opening angle assumes a conical outflow, in contrast with the frustrumal geometry assumed with our previous photometric estimate. To more directly compare the two methods, we derive an opening angle from our photometric results assuming a conical geometry. We fit a linear relationship to the surface-brightness derived outflow region as above, but enforce a y-intercept of 0, resulting in an outflow profile that terminates at the galaxy center. By design this results in a larger opening angle, since the position of the 50\% contour at large z-height has not changed, but the origin has. With this method we calculate an opening angle of $\sim$37$^{\circ}$ in the NW region and $\sim$32$^{\circ}$ in the SE region, which are significantly larger than and roughly twice as large as the angles calculated with either previous method. At the 90\% width, we measure an opening angle of $\sim$73$^{\circ}$ in the NW region and $\sim$74$^{\circ}$ in the SE region. These values are again significantly larger than their frustrumal counterparts.

It is interesting to note that the surface brightness method identifies a similar opening angle as the line splitting, despite the significant number of differences in measurements and assumptions. However, this agreement relies on an assumption of $v_{\rm out}\sim300$~km~s$^{-1}$. If we change this to a lower value of 150~km~s$^{-1}$, which is still within the range of $v_{\rm out}$ for this SFR and mass, then the line-splitting based opening angle changes to 34$^{\circ}$. This illustrates the challenge of estimating the opening angle using the line splitting, as it is heavily based on the (very uncertain) adopted $v_{\rm out}$. 

\begin{table}
	\centering
	\caption{Calculations of the minor-axis outflow opening angle with different methods and assumptions as discussed in Section~\ref{subsec:idoutflow}.}
	\label{tab:outwidths}
	\begin{tabular}{cccc} 
		\hline
		Method & Assumed Shape & Region & $\theta$ ($^{\circ}$)\\
		\hline
            Photometry & Frustrum & NW & 19.81 $\pm$ 0.01\\
            Photometry & Frustrum & SE & 17.70 $\pm$ 0.02\\
            Line Splitting & Conical & NW & 16.96 $\pm$ 0.41\\
            Line Splitting & Conical & SE & 17.27 $\pm$ 0.14\\
            Photometry & Conical & NW & 37.37 $\pm$ 0.01\\
            Photometry & Conical & SE & 31.68 $\pm$ 0.01\\
		\hline
	\end{tabular}
\end{table}

\subsection{Outflows in the plane of the disk}
\label{subsec:diskout}
\begin{figure*}
    \centering
    \includegraphics[width=\textwidth]{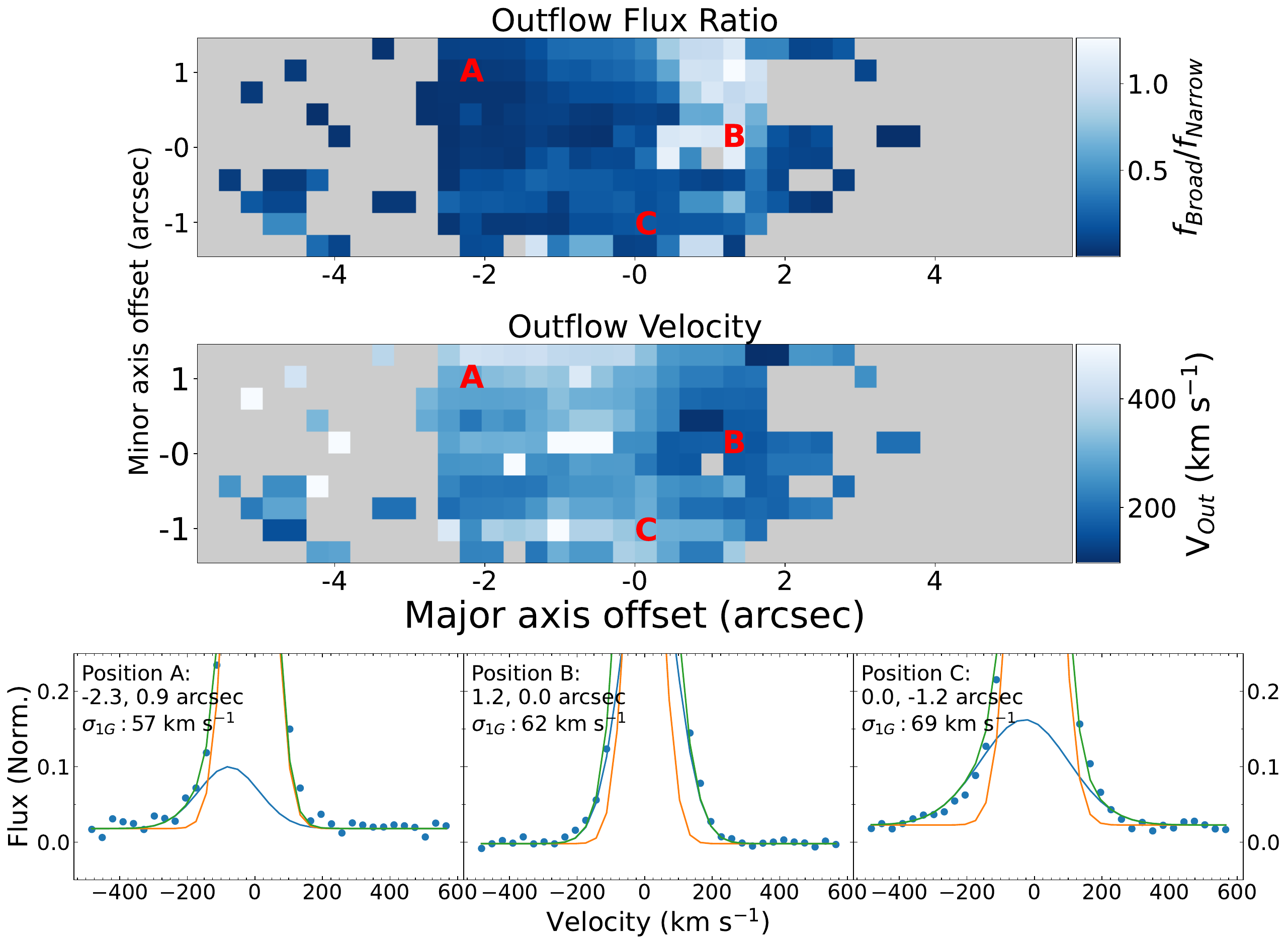}
    \caption{Map across the disk edge of \emph{Top:} ratio of the broad component flux to the narrow component flux in spaxels where the secondary component is detected, \emph{Middle:} outflow velocity calculated for spaxels where the secondary component is detected, and \emph{Bottom:} representative fits in positions A, B, and C as shown in the upper and middle panels. The flux ratio and outflow velocity are only calculated in spaxels where there is statistically significant evidence for multiple components based on the BIC value.}
    \label{fig:diskmassout}
\end{figure*}
Using the line fitting method described in Section~\ref{subsec:linefit} we find multiple components are common within the plane of the disk. This is seen in the bottom panel of Fig.~\ref{fig:line_splitting} in which the secondary components make up significant fractions of the total flux inside $\sim \pm1$~kpc of the disk midplane. This could suggest gas outflows along the line of sight, especially for those that are blueshifted. We find that in the vast majority of spaxels ($\sim$80-90\%) the broad component is blueshifted with respect to the narrow emission line, which is consistent with expectations for outflows.  Similarly, \citet{Chisholm2015} measures the outflow via decomposition of the absorption line, which would be in the plane of the disk for this edge-on galaxy.  Here we consider these in a similar way as is done in face-on galaxies \citep[e.g][]{ReichardtChu2022}.

By decomposing the observed emission lines in the disk region into a broad outflow component and a narrow systemic component as described above, and then taking the total broad and narrow flux across the disk region we measure the ratio of the outflow to systemic flux, $F_{\rm broad}/F_{\rm narrow}=0.21$. We show a map of this quantity in the top panel of fig.~\ref{fig:diskmassout}. This is consistent with results from \citet{Davies2019} and \citet{ReichardtChu2022}. At higher redshift, \citet{Concas2022} argues that outflows from similar mass galaxies as Mrk~1486 may be overestimated due to spatial resolution effects in KMOS observations \citep[e.g.][]{Forster2019}. Our spatial resolution is significantly higher than those observations. We find, even in the edge-on case, that there is significant broadline flux, similar to studies like \citet{Forster2019}.
If the secondary component in the plane of the disk is truly outflow, in principle one assumes it would be a larger fraction of the flux for a face-on galaxy. 

We can estimate the mass loading in the plane of the disk, albeit with significant systematic uncertainty introduced by necessary assumptions. Taking the broad component to comprise the outflow we measure the resolved mass outflow rate from the disk, along the line-of-sight via:
\begin{equation}
\label{eq:moutdownthebarrel}
    \dot{M}_{\rm out}=\frac{1.36m_{H}}{\gamma_{H\beta}n_{e}}\frac{v_{\rm out}}{R_{\rm out}}L_{\rm H\beta, broad},
\end{equation}
where all constants are as described in previous equations, $L_{\rm H\beta, broad}$ is the luminosity in the broad component of the H$\beta$ emission line, and $R_{\rm out}$ is the radial extent of the outflow region. In Eq.~\ref{eq:moutdownthebarrel} the two largest sources of uncertainty are the outflow radius, $R_{\rm out}$ and the electron density, $n_{e}$. For electron density, we adopt a single value of $n_{e}=32$~cm$^{-3}$ based on our measurement of this quantity in the disk in Section~\ref{subsec:elecdensity}. If the electron density in this outflow is however decaying with radial position, our derived value for $\dot{M}_{\rm out}$ will then be a lower bound for a given outflow velocity and radius ($R_{\rm out}$). $R_{\rm out}$ is not directly measurable in down-the-barrel observations. As the morphology of this edge-on outflow is not well known, and there is little precedent in the literature, this introduces a large systematic uncertainty into calculations of $\dot{M}_{\rm out}$. Additionally, the value of $R_{\rm out}$ may vary from spaxel to spaxel, as the $\dot{M}_{\rm out}$ may include gas originating from within a given spaxel as well as tangentially launched gas from adjacent spaxels. We measure a maximum full-width of 2kpc at $1$~kpc minor axis offset for the $50\%$ width of the minor axis outflow region, which provides a possible upper limit for $R_{\rm out}$ in the edge-on outflow. Both the size and electron density are consistent with values derived from modeling of UV absorption lines from \citet{xu2023}. We assume a constant value of $R_{\rm out}$ across all outflow spaxels. For the assumption of $R_{\rm out}=2$~kpc and $n_e=32$~cm$^{-3}$ we find $\dot{M}_{\rm out}({\rm edge})=0.9$~M$_{\odot}$~yr$^{-1}$, which yields a mass-loading of $\eta({\rm edge}) \sim 0.25$. We will show in subsequent sections that this is of order 25\% of the outflows above the plane of the disk. 

In Fig.~\ref{fig:diskmassout} there is a region at position $\sim$1.75 arcsec along the major axis and $\sim$1~arcsec above the plane on the minor axis that has  $f_{\rm broad}/f_{\rm narrow}\sim1$. This area is also seen in Fig.~\ref{fig:line_splitting} and sits near the NW minor axis outflow lobe. The position is elevated above the midplane, and the boundary between outflow and disk may not be well defined at all places in the galaxy. It is, therefore, plausible that this region is not truly an outflow in the plane of the disk, but rather the beginning of the minor axis outflow. A 1.2$\times$1.2~arcsec$^2$ region centered on the peak $f_{\rm broad}/f_{\rm narrow}$ in this area comprises $\sim10$\% of the outflow flux in the plane of the disk. 

Unlike in the extraplanar regions directly above and below the disk, with kinematically decomposed observations of the outflow in the disk region we can directly measure the outflow velocity. Within this region, the outflow velocity is defined as:
\begin{equation}
    v_{\rm out}=|\Delta\mu| + 2\sigma_{\rm broad}.
\end{equation}
Here $\Delta\mu$ is the difference between the means of the systemic and outflow Gaussians and $\sigma_{\rm broad}$ is the spread in the broad Gaussian component. A map of $v_{\rm out}$ across the disk region of Mrk~1486 is shown in the middle panel of Fig.~\ref{fig:diskmassout}. We note that in the figure, and in our analysis, we only show spaxels with statistically significant evidence for multiple components based on the BIC values. Within this region we measure outflow velocities within the range $100-430$~km~s$^{-1}$, with a median value of $220$~km~s$^{-1}$. These measured velocities are consistent with the range of velocities considered in the minor axis mass outflow rate profile and with those described in the literature \citep{Chisholm2015, Heckman2015}.

\subsection{Outflows along the minor axis} 
\label{subsec:minoraxisoutflow}

\begin{figure}
    \centering
    \includegraphics[width=\linewidth]{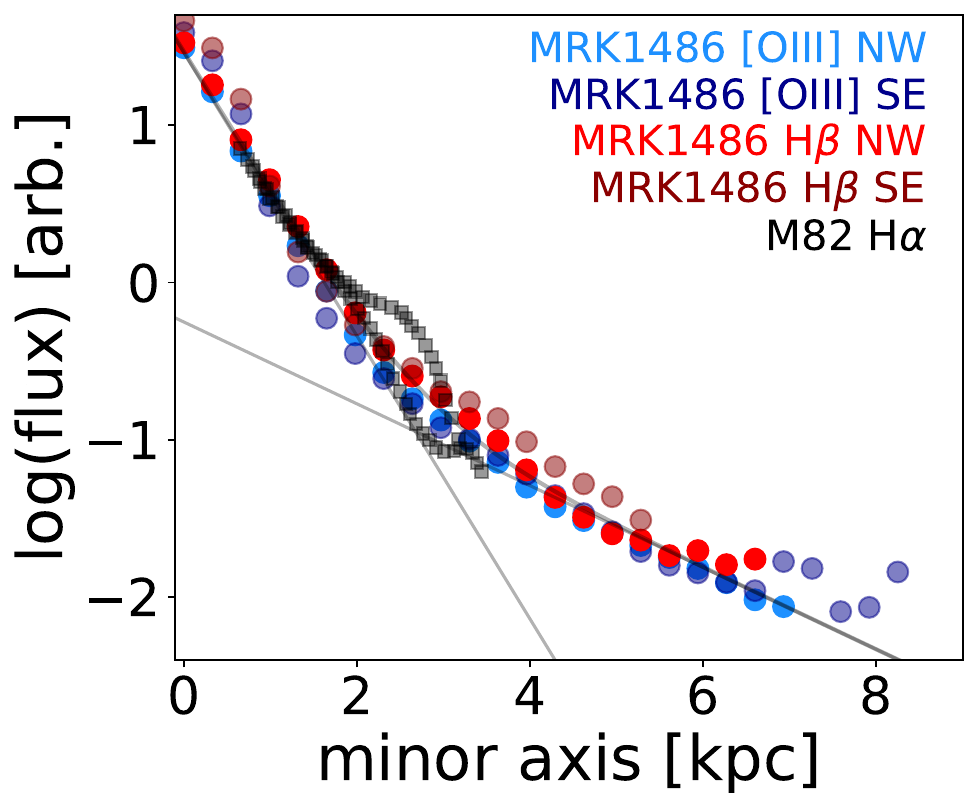}
    \caption{Radial flux profiles for the minor-axis outflow regions in Mrk~1486. Shown are profiles for [OIII]~$\lambda$5007 and H$\beta$ in both the NW and SE outflow regions. The two outflow regions are very symmetric in terms of surface brightness. Overlaid in grey is the same profile for H$\alpha$ emission in M~82 from \citet{Leroy2015}. Overplotted is the average double exponential fit to the shown flux profiles in Mrk~1486, and the individual exponential components, showing the galaxy and outflow exponential. In the regions for which the M82 data is available, the shape of the profile is very similar to Mrk~1486.} 
    \label{fig:surfacebrightness}
\end{figure}

Fig.~\ref{fig:surfacebrightness} shows the vertical axis surface brightness profiles in [OIII] and H$\beta$, obtained by taking the mean at each minor axis offset within the outflow region determined via \textsc{threadcount}. We recover the extended emission profile consistent with minor axis outflows. We note that the surface brightness profile is a smooth decay with radius that does not support a ``thin-shell" geometry, which is sometimes invoked in the literature to interpret outflows measured via quasar absorption lines \citep[see discussion in][]{Veilleux2020, Tumlinson2017}. 

While the two minor axis outflow regions are highly asymmetric in their morphological and kinematic properties, the shape of the surface brightness profile is remarkably similar between the two. We fit a double exponential to the surface brightness profile and treat the outer exponential as the outflow profile. The average fit to these profiles is overplotted in fig.~\ref{fig:surfacebrightness}, as are the component single exponential profiles. Within this outflow profile we find a scale height of $2.5$~kpc in the NW outflow region and $2.1$~kpc in the SE outflow region. This implies that the majority of the outflow mass is contained within $\sim$3-4~kpc of disk center. We measure a $90\%$ extent (of the total integrated flux) of $4.7$~kpc in the upper outflow and $5.7$~kpc in the lower outflow. Some studies use the maximum observed position of the outflow as the ``outflow extent". This is clearly dependent on the sensitivity of the observations. The $90\%$ extent (or even the 50\% distance) could be considered an analogous quantity and provides a direct observable that can be compared to simulation results that is less impacted by sensitivity. 

With the minor axis outflow region determined we can consider the total mass outflow rate within this region. For a minor axis outflow comprised of $n$ strips of spaxels perpendicular to the galaxy disk, the total mass outflow rate within this outflow region is given by:
\begin{equation}
    \dot{M}_{\rm out}=\sum_{i=0}^{n} \frac{v_{{\rm out}, i}}{r_{i}}\frac{1.36 m_{H}}{n_{e, i}\gamma_{\rm H\beta}}L_{{\rm H\beta}, i},
\end{equation}
based on the thick wind formulae from \citet{Rupke2005}. Here $L_{{\rm H\beta}, i}$ is the total H$\beta$ luminosity within strip $i$, $m_{H}$ is the atomic mass of Hydrogen, $n_{e,i}$ is the electron density in strip $i$, $\gamma_{\rm H\beta}$ is the H$\beta$ emissivity, $v_{{\rm out}, i}$ is the outflow velocity within strip $i$, and $r_{i}$ is the distance between strip $i$ and the galaxy. From the [OIII]~$\lambda4363$ emission line, the temperature in the outflow is determined to be $\sim1.2-1.3\times10^{4}$~K \citep{Cameron2021}. We expect the [OIII] to be biased towards hotter gas compared with H$\beta$ and adopt a temperature $T=10^{4}$~K, resulting in $\gamma_{\rm H\beta}=1.24\times10^{-25}$~erg~cm$^{3}$~s$^{-1}$. The two largest sources of uncertainty in this determination of $\dot{M}_{\rm out}$ are the gradients in the electron density and in the velocity of the wind. These are discussed at length in Appendix~\ref{sec:uncertainparams}. The electron density in particular introduces a large, systematic uncertainty into any determination of the mass outflow rate in this galaxy, and within the outflow region is entirely unconstrained. As Mrk~1486 and its outflow are at densities $\leq 32$~cm$^{-3}$ all spaxels within our data are at or below the low-density limit for all optical emission line tracers of density. In our calculation of the total mass outflow rate within these minor axis lobes, we assume a decaying electron density profile following:
\begin{equation}
\label{eq:electrondensityneg1}
    n_{e, z}=n_{e, {\rm max}}\left(\frac{z}{h_{n_{e}}}\right)^{-1}
\end{equation}
with an electron density scale height of $h_{n_{e}}=0.8$~kpc, such that the density reaches the maximum value at the disk edge. As Mrk~1486 is at an inclination of 85$^\circ$, we cannot observe the radial outflow velocity in the minor axis regions and must make assumptions about the shape of the velocity profile in the outflow. There is precedent in the literature for a maximum outflow velocity between 150 and 450~km~s$^{-1}$ \citep{Chisholm2015, Heckman2015}. Appendix~\ref{subsec:appendixvelocity} details our determination of the impact of the outflow velocity profile on the final calculated mass outflow rate. We find that with reasonable assumptions the choice of velocity profile has a minimal impact on the mass outflow rate in the edge on outflow. We thus assume a constant outflow velocity with $v_{\rm out}=300$~km~s$^{-1}$. With the above assumptions, we calculate a mass outflow rate at the 90\% width of 1.2~M$_{\odot}$~yr$^{-1}$ in the NW outflow region and 1.3~M$_{\odot}$~yr$^{-1}$ in the SE region. This corresponds to a total mass outflow rate in the minor axis regions of $\dot{M}_{\rm out}({\rm minor})=2.5$~M$_{\odot}$~yr$^{-1}$ and a mass loading factor $\eta({\rm minor})=0.7$.

\subsection{3-Dimensional Outflow Shape}
\label{subsec:3dmassout}

\begin{figure}
    \centering
    \includegraphics[width=\linewidth]{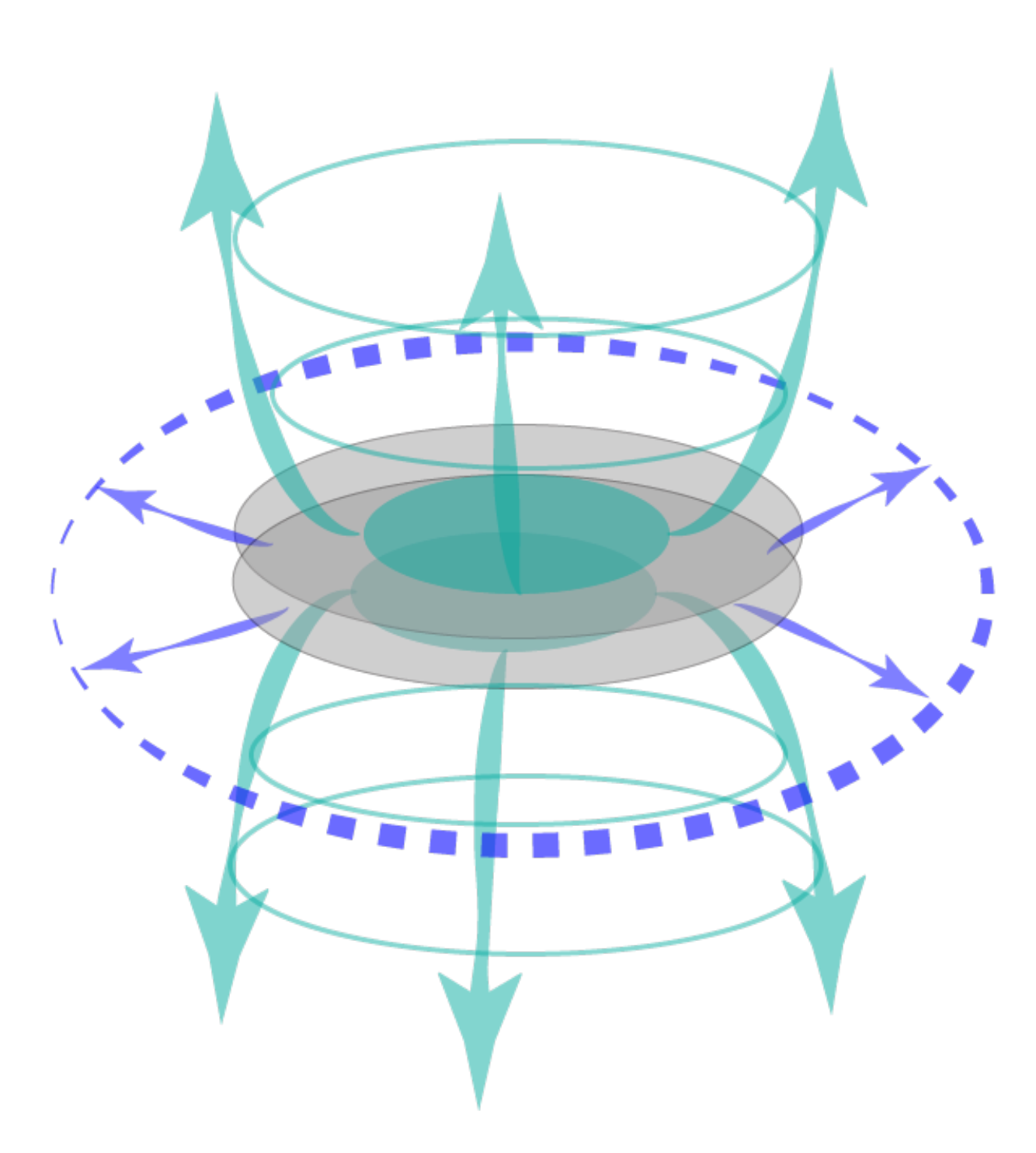}
    \caption{A cartoon representation of the 3D outflow shape in Mrk~1486. The galaxy disk is shown in grey. We detect a bifrustrumal outflow component along the minor axis (teal) that dominates the mass outflow in the galaxy. In addition, we detect a disk-edge component (blue) containing a non-negligible ($\sim25\%$) amount of the total outflowing mass. The 3D-shape of the outflow in Mrk~1486 is thus not purely spherical nor only biconical. If this outflow shape is common, then this suggests systematic uncertainties in the calculation of the total mass outflow rate in galaxies where only the minor axis or disk-edge outflow is detected.}
    \label{fig:3d_outflow_shape}
\end{figure}

In the above sections, we have estimated the mass outflow rate in both the disk-edge and minor axis outflow regions. We show in Fig.~\ref{fig:3d_outflow_shape} a cartoon diagram that is representative of the outflow shape for Mrk~1486. The minor axis outflow, which dominates in terms of mass flux, is shown in cyan and the disk-edge outflow, which contributes $\sim25\%$ to the total outflow mass rate in the galaxy, is shown in blue. We determine the outflow shape in Mrk~1486 to be neither purely spherical nor biconical. The outflow is still dominated by the bifrustrumal shape that is typical of other outflows studied \citep[e.g.][]{Shopbell1998,Herenz2023}, but has another lower azimuth component. 

If this complex outflow shape is not unique to Mrk~1486, then the relative contributions of the different outflow components suggest possible systematic uncertainties in the determination of the total mass outflow rate for a galaxy observed with measurements that only detect a subset of the components. We now consider the impact of this systematic. To do so we compare the mass outflow rate of each component to that of a thin, spherical shell of similar flux. This simulates observation of each component were it an unresolved, ``down-the-barrel" observation. 

For each outflow component we determine a total mass and the area in which outflows are detected. We then calculate a mass density $\rho_{\rm outflow}$ within the determined outflow region and let $\rho_{\rm outflow} = \mu m_{p} N_{H}$ in Eq.~\ref{eq:mout}. We then assume $\Omega=4\pi$ per the spherical geometry.  We adopt $R_{\rm out} = 2$~kpc (as in our determination of the disk-edge mass outflow rate in Section~\ref{subsec:diskout}), and $v_{\rm out}=300$~km~s$^{-1}$ (as in our determination of the minor-axis outflow rate in Section~\ref{subsec:minoraxisoutflow} and consistent with the range of observed velocities in the disk edge outflow). We then calculate a mass outflow rate assuming a spherical outflow morphology via. Eq/~\ref{eq:mout}. 

Table~\ref{tab:3d_shape} shows the calculated mass outflow rate based on whether the disk-edge or minor axis outflows are observed. Note that the middle column, ``Empirical Shape,'' is not the global mass-loading for Mrk~1486, but only the mass rate of that component. For the entire galaxy we measure a total empirical mass outflow rate of $\dot{M}_{\rm out}=3.4$~M$_{\odot}$~yr$^{-1}$. On the minor-axis the outflow is extended over a large area, and thus lowering the surface density of any single line of sight that is observed. As the spherical, thin-shell geometry neglects the large extent of the minor axis outflows, their low surface density results in low total mass outflow rates of $\dot{M}_{\rm out}=1.6$~M$_{\odot}$~yr$^{-1}$ whether the mass density is calculated from the NW or SE lobe. Conversely, the relative high surface density of the disk-edge outflow results in a very large total mass outflow rate of $\dot{M}_{\rm out}=9.2$~M$_{\odot}$~yr$^{-1}$, much greater than the empirically determined value. Clearly, the geometry of the galaxy, and thus outflow, is an important parameter to constrain in low-resolution observations of outflows \citep[e.g.][]{Rubin2014,Forster2019}.

\begin{table}
	\centering
	\caption{The mass outflow rate within each detected outflow component and the calculated mass outflow rate assuming a thin-shell, spherical outflow morphology as discussed in Section~\ref{subsec:3dmassout}.}
	\label{tab:3d_shape}
	\begin{tabular}{ccc} 
		\hline
		Observed Component & Empirical Shape & Spherical Model\\
            & $(M_{\odot}$~yr$^{-1})$ & $(M_{\odot}$~yr$^{-1})$\\
		\hline
            Minor Axis NW & 1.2 & 1.6\\
		  Minor Axis SE & 1.3 & 1.6\\
		  Disk-Edge & 0.9 & 9.2\\
		\hline
	\end{tabular}
\end{table}

\section{Discussion \& Conclusion} \label{sec:discussion}

In this paper we analyzed the extraplanar emission in H$\beta$ and [OIII]~$\lambda5007$ of the galaxy Mrk~1486. We present a simple technique for identifying outflows in edge-on galaxies and measuring their properties, which can be applied to IFU observations or narrowband imaging with sufficient spatial resolution to resolve the outflow. The outflow emission is identified by (1) identifying lines perpendicular to the galaxy disk which correspond to the peak surface brightness above and below the disk (ie the outflow axis) and (2) measuring the surface brightness contours that enclose the width of the outflow. In this work we chose the 50\% or 90\% widths, however this could be altered in future work based on the needs of the program. In Fig.~\ref{fig:line_splitting} we show that this method, which only requires emission line surface brightness, recovers the same region that is identified as having larger line-widths and stronger presence of multiple components. We take this as confirmation that the simple photometry is identifying a kinematically-distinct component of the extraplanar material, consistent with an expanding outflow. 

We observe emission to $\pm$8~kpc above and below the disk, far beyond the distance of stellar continuum. The outflow follows an exponential decay with scale-length of $\sim$2.1-2.5~kpc, and corresponding distance that encloses 90\% of the outflow flux to be $\sim$5-6~kpc. The outflow has a wide base, which reflects the wide distribution of star formation in Mrk~1486. This base is wider than that measured in M82. 

In addition, we observe a disk-edge outflow component, contributing $\sim25\%$ to the total mass outflow rate for the galaxy. We are thus able to determine the 3-dimensional shape of the outflow in Mrk~1486, and find it to be neither spherical nor purely biconical. A plausible cause of both the wide-base and disk-plane outflows is the distribution of star formation within Mrk~1486. We find strong emission lines extending to the edge of the disk, and likewise wide-spread UV emission. This is indicative of a galaxy-wide starburst. The SFR of Mrk~1486 is roughly 5$\times$ higher than the main-sequence for its mass, which likewise indicates a strong amount of star formation. If we compare Mrk~1486 to the local, well-studied outflow galaxy NGC~253 there are significant differences in the global distribution of star formation. In NGC~253 the starburst is contained within a small region in the central $R\sim500$~pc of a large disk. There are many kiloparsecs between the starburst and the disk edge of NGC~253, which would act to collimate the outflow. In Mrk~1486 the starburst is not centrally located, which allows for escape from the disk along more lines of sight. We note this may have implications for outflows from galaxies at larger redshift, which do not necessarily have a centrally concentrated star formation and may be clumpier. 

The mass distribution within these outflow regions is heavily dependent on assumptions about electron density and $R_{\rm out}$. Under the assumption of a constant outflow velocity and a radial electron density profile decaying with $r^{-1}$, we measured a total mass outflow rate within the minor axis outflow of $2.5$~M$_{\odot}$~yr$^{-1}$. The electron density is essentially unconstrained in the extraplanar regions. The entirety of the galaxy and especially the extraplanar regions are at or below the low-density limit for all optical emission line tracers of density. While observations with greater spectral resolution would assist in resolving the individual [OII] emission lines and observations of the [SII] doublet as another tracer of electron density that is more widely separated and thus more easily resolved, the determination of the electron density within the outflow regions is still hampered by the limitations of optical emission line tracers of density. 

We calculated the opening angle for the minor axis outflow using multiple methods and assumptions. In particular, we compared the opening angle determined in the typical manner, by measuring the degree of separation between components in multi-Gaussian fits to the observed emission lines, with the opening angle measured directly from our determination of the outflow region. Our photometric determination of the outflow region produces an opening angle at the $50\%$ width of $20^{\circ}$ in the NW region and $18^{\circ}$ in the SE region assuming a frustrumal outflow shape, and $37^{\circ}$ in the NW region and $32^{\circ}$ in the SE region assuming a conical outflow shape. Using the more common ``line-splitting'' technique for determining the opening angle we calculate an opening angle of $17^{\circ}$ in both the NW and SE regions. This opening angle determination is consistent with the opening angle determined from the \textsc{threadcount} outflow shape, supporting this method of direct opening angle observation.  Determination of the opening angle and geometry of galactic winds opens an interesting new discovery for studying outflows, which will facilitate much more precise determinations of the mass outflow rates of galactic outflows.

The sample size of galaxies for which we have resolved outflow observations is very small, and thus whenever we can add another target to that sample, the results on that target are worth noting as we have done here with Mrk~1486. This galaxy has similar ISM characteristics to galaxies at larger redshift in that it is low metallicity, high SFR, and low mass. An option to reproduce these results at higher redshift is to use long integrations with JWST NIRSpec PRISM, which has extremely high throughput. While this would sacrifice kinematic information and likely blend emission lines, one could generate a map of ionized gas in visible wavelengths for comparison to simulations. Our results present a new method, albeit representing only one object. More observations of outflowing systems are required. The method developed above can be easily applied to larger samples of galaxies observed with KCWI and MUSE to map resolved gas flows and their physical properties. 

\section*{Acknowledgements}

Parts of this research were supported by the Australian Research Council Centre of Excellence for All Sky Astrophysics in 3 Dimensions (ASTRO 3D), through project number CE170100013.  
D.B.F. acknowledges support from Australian Research Council (ARC) Future Fellowship FT170100376 and ARC Discovery Program grant DP130101460. 
A.D.B. acknowledges partial support from AST1412419 and AST2108140. 
A.J.C. acknowledges funding from the ``FirstGalaxies'' Advanced Grant from the European Research Council (ERC) under the European Union's Horizon 2020 research and innovation programme (Grant agreement No. 789056).
R.H.-C. thanks the Max Planck Society for support under the Partner Group project "The Baryon Cycle in Galaxies" between the Max Planck for Extraterrestrial Physics and the Universidad de Concepción. R.H-C also acknowledges financial support from Millennium Nucleus NCN19058 (TITANs) and support by the ANID BASAL projects ACE210002 and FB210003.
R.R.V. and K.S. acknowledge funding support from National Science Foundation Award No. 1816462.

Some of the data presented herein were obtained at the W. M. Keck Observatory, which is operated as a scientific partnership among the California Institute of Technology, the University of California and the National Aeronautics and Space Administration. The Observatory was made possible by the generous financial support of the W. M. Keck Foundation. Observations were supported by Swinburne Keck. The authors wish to recognise and acknowledge the very significant cultural role and reverence that the summit of Maunakea has always had within the indigenous Hawaiian community. We are most fortunate to have the opportunity to conduct observations from this mountain.
\section*{Data Availability}

The DUVET Survey is still in progress. The data underlying this article will be shared on reasonable request to the PI, Deanne Fisher at dfisher@swin.edu.au



\bibliographystyle{mnras}
\bibliography{bibliography} 

\begin{thebibliography}{}
\makeatletter
\relax
\def\mn@urlcharsother{\let\do\@makeother \do\$\do\&\do\#\do\^\do\_\do\%\do\~}
\def\mn@doi{\begingroup\mn@urlcharsother \@ifnextchar [ {\mn@doi@}
  {\mn@doi@[]}}
\def\mn@doi@[#1]#2{\def\@tempa{#1}\ifx\@tempa\@empty \href
  {http://dx.doi.org/#2} {doi:#2}\else \href {http://dx.doi.org/#2} {#1}\fi
  \endgroup}
\def\mn@eprint#1#2{\mn@eprint@#1:#2::\@nil}
\def\mn@eprint@arXiv#1{\href {http://arxiv.org/abs/#1} {{\tt arXiv:#1}}}
\def\mn@eprint@dblp#1{\href {http://dblp.uni-trier.de/rec/bibtex/#1.xml}
  {dblp:#1}}
\def\mn@eprint@#1:#2:#3:#4\@nil{\def\@tempa {#1}\def\@tempb {#2}\def\@tempc
  {#3}\ifx \@tempc \@empty \let \@tempc \@tempb \let \@tempb \@tempa \fi \ifx
  \@tempb \@empty \def\@tempb {arXiv}\fi \@ifundefined
  {mn@eprint@\@tempb}{\@tempb:\@tempc}{\expandafter \expandafter \csname
  mn@eprint@\@tempb\endcsname \expandafter{\@tempc}}}

\bibitem[\protect\citeauthoryear{{Bolatto} et~al.,}{{Bolatto}
  et~al.}{2013}]{Bolatto2013}
{Bolatto} A.~D.,  et~al., 2013, \mn@doi [\nat] {10.1038/nature12351}, \href
  {https://ui.adsabs.harvard.edu/abs/2013Natur.499..450B} {499, 450}

\bibitem[\protect\citeauthoryear{{Burchett}, {Rubin}, {Prochaska}, {Coil},
  {Vaught}  \& {Hennawi}}{{Burchett} et~al.}{2021}]{Burchett2021}
{Burchett} J.~N.,  {Rubin} K. H.~R.,  {Prochaska} J.~X.,  {Coil} A.~L.,
  {Vaught} R.~R.,   {Hennawi} J.~F.,  2021, \mn@doi [\apj]
  {10.3847/1538-4357/abd4e0}, \href
  {https://ui.adsabs.harvard.edu/abs/2021ApJ...909..151B} {909, 151}

\bibitem[\protect\citeauthoryear{{Calzetti}}{{Calzetti}}{2001}]{Calzetti2001}
{Calzetti} D.,  2001, \mn@doi [\pasp] {10.1086/324269}, \href
  {https://ui.adsabs.harvard.edu/abs/2001PASP..113.1449C} {113, 1449}

\bibitem[\protect\citeauthoryear{{Cameron} et~al.,}{{Cameron}
  et~al.}{2021}]{Cameron2021}
{Cameron} A.~J.,  et~al., 2021, \mn@doi [\apjl] {10.3847/2041-8213/ac18ca},
  \href {https://ui.adsabs.harvard.edu/abs/2021ApJ...918L..16C} {918, L16}

\bibitem[\protect\citeauthoryear{{Cappellari}}{{Cappellari}}{2017}]{Cappellari2017}
{Cappellari} M.,  2017, \mn@doi [\mnras] {10.1093/mnras/stw3020}, \href
  {https://ui.adsabs.harvard.edu/abs/2017MNRAS.466..798C} {466, 798}

\bibitem[\protect\citeauthoryear{{Chevalier} \& {Clegg}}{{Chevalier} \&
  {Clegg}}{1985}]{Chevalier1985}
{Chevalier} R.~A.,  {Clegg} A.~W.,  1985, \mn@doi [\nat] {10.1038/317044a0},
  \href {https://ui.adsabs.harvard.edu/abs/1985Natur.317...44C} {317, 44}

\bibitem[\protect\citeauthoryear{{Chisholm}, {Tremonti}, {Leitherer}, {Chen},
  {Wofford}  \& {Lundgren}}{{Chisholm} et~al.}{2015}]{Chisholm2015}
{Chisholm} J.,  {Tremonti} C.~A.,  {Leitherer} C.,  {Chen} Y.,  {Wofford} A.,
  {Lundgren} B.,  2015, \mn@doi [\apj] {10.1088/0004-637X/811/2/149}, \href
  {https://ui.adsabs.harvard.edu/abs/2015ApJ...811..149C} {811, 149}

\bibitem[\protect\citeauthoryear{{Chisholm}, {Bordoloi}, {Rigby}  \&
  {Bayliss}}{{Chisholm} et~al.}{2018a}]{Chisholm2018}
{Chisholm} J.,  {Bordoloi} R.,  {Rigby} J.~R.,   {Bayliss} M.,  2018a, \mn@doi
  [\mnras] {10.1093/mnras/stx2848}, \href
  {https://ui.adsabs.harvard.edu/abs/2018MNRAS.474.1688C} {474, 1688}

\bibitem[\protect\citeauthoryear{{Chisholm}, {Tremonti}  \&
  {Leitherer}}{{Chisholm} et~al.}{2018b}]{Chisholm2018b}
{Chisholm} J.,  {Tremonti} C.,   {Leitherer} C.,  2018b, \mn@doi [\mnras]
  {10.1093/mnras/sty2380}, \href
  {https://ui.adsabs.harvard.edu/abs/2018MNRAS.481.1690C} {481, 1690}

\bibitem[\protect\citeauthoryear{{Concas} et~al.,}{{Concas}
  et~al.}{2022}]{Concas2022}
{Concas} A.,  et~al., 2022, \mn@doi [\mnras] {10.1093/mnras/stac1026}, \href
  {https://ui.adsabs.harvard.edu/abs/2022MNRAS.513.2535C} {513, 2535}

\bibitem[\protect\citeauthoryear{{Davies} et~al.,}{{Davies}
  et~al.}{2019}]{Davies2019}
{Davies} R.~L.,  et~al., 2019, \mn@doi [\apj] {10.3847/1538-4357/ab06f1}, \href
  {https://ui.adsabs.harvard.edu/abs/2019ApJ...873..122D} {873, 122}

\bibitem[\protect\citeauthoryear{{Duval} et~al.,}{{Duval}
  et~al.}{2016}]{Duval2016}
{Duval} F.,  et~al., 2016, \mn@doi [\aap] {10.1051/0004-6361/201526876}, \href
  {https://ui.adsabs.harvard.edu/abs/2016A&A...587A..77D} {587, A77}

\bibitem[\protect\citeauthoryear{{Fielding} \& {Bryan}}{{Fielding} \&
  {Bryan}}{2022}]{Fielding2022}
{Fielding} D.~B.,  {Bryan} G.~L.,  2022, \mn@doi [\apj]
  {10.3847/1538-4357/ac2f41}, \href
  {https://ui.adsabs.harvard.edu/abs/2022ApJ...924...82F} {924, 82}

\bibitem[\protect\citeauthoryear{{F{\"o}rster Schreiber} et~al.,}{{F{\"o}rster
  Schreiber} et~al.}{2019}]{Forster2019}
{F{\"o}rster Schreiber} N.~M.,  et~al., 2019, \mn@doi [\apj]
  {10.3847/1538-4357/ab0ca2}, \href
  {https://ui.adsabs.harvard.edu/abs/2019ApJ...875...21F} {875, 21}

\bibitem[\protect\citeauthoryear{{Hayward} \& {Hopkins}}{{Hayward} \&
  {Hopkins}}{2017}]{Hayward2017}
{Hayward} C.~C.,  {Hopkins} P.~F.,  2017, \mn@doi [\mnras]
  {10.1093/mnras/stw2888}, \href
  {https://ui.adsabs.harvard.edu/abs/2017MNRAS.465.1682H} {465, 1682}

\bibitem[\protect\citeauthoryear{{Heckman}, {Alexandroff}, {Borthakur},
  {Overzier}  \& {Leitherer}}{{Heckman} et~al.}{2015}]{Heckman2015}
{Heckman} T.~M.,  {Alexandroff} R.~M.,  {Borthakur} S.,  {Overzier} R.,
  {Leitherer} C.,  2015, \mn@doi [\apj] {10.1088/0004-637X/809/2/147}, \href
  {https://ui.adsabs.harvard.edu/abs/2015ApJ...809..147H} {809, 147}

\bibitem[\protect\citeauthoryear{{Herenz} et~al.,}{{Herenz}
  et~al.}{2023}]{Herenz2023}
{Herenz} E.~C.,  et~al., 2023, \mn@doi [\aap] {10.1051/0004-6361/202244930},
  \href {https://ui.adsabs.harvard.edu/abs/2023A&A...670A.121H} {670, A121}

\bibitem[\protect\citeauthoryear{{Ho} et~al.,}{{Ho} et~al.}{2016}]{Ho2016}
{Ho} I.~T.,  et~al., 2016, \mn@doi [\mnras] {10.1093/mnras/stw017}, \href
  {https://ui.adsabs.harvard.edu/abs/2016MNRAS.457.1257H} {457, 1257}

\bibitem[\protect\citeauthoryear{{Leroy} et~al.,}{{Leroy}
  et~al.}{2015}]{Leroy2015}
{Leroy} A.~K.,  et~al., 2015, \mn@doi [\apj] {10.1088/0004-637X/814/2/83},
  \href {https://ui.adsabs.harvard.edu/abs/2015ApJ...814...83L} {814, 83}

\bibitem[\protect\citeauthoryear{{Lopez}, {Mathur}, {Nguyen}, {Thompson}  \&
  {Olivier}}{{Lopez} et~al.}{2020}]{Lopez2020}
{Lopez} L.~A.,  {Mathur} S.,  {Nguyen} D.~D.,  {Thompson} T.~A.,   {Olivier}
  G.~M.,  2020, \mn@doi [\apj] {10.3847/1538-4357/abc010}, \href
  {https://ui.adsabs.harvard.edu/abs/2020ApJ...904..152L} {904, 152}

\bibitem[\protect\citeauthoryear{{Marasco} et~al.,}{{Marasco}
  et~al.}{2022}]{Marasco2022}
{Marasco} A.,  et~al., 2022, arXiv e-prints, \href
  {https://ui.adsabs.harvard.edu/abs/2022arXiv220902726M} {p. arXiv:2209.02726}

\bibitem[\protect\citeauthoryear{{Martin}}{{Martin}}{2005}]{Martin2005}
{Martin} C.~L.,  2005, \mn@doi [\apj] {10.1086/427277}, \href
  {https://ui.adsabs.harvard.edu/abs/2005ApJ...621..227M} {621, 227}

\bibitem[\protect\citeauthoryear{{Morrissey} et~al.,}{{Morrissey}
  et~al.}{2018}]{Morrissey2018}
{Morrissey} P.,  et~al., 2018, \mn@doi [\apj] {10.3847/1538-4357/aad597}, \href
  {https://ui.adsabs.harvard.edu/abs/2018ApJ...864...93M} {864, 93}

\bibitem[\protect\citeauthoryear{{Naab} \& {Ostriker}}{{Naab} \&
  {Ostriker}}{2017}]{Naab2017}
{Naab} T.,  {Ostriker} J.~P.,  2017, \mn@doi [\araa]
  {10.1146/annurev-astro-081913-040019}, \href
  {https://ui.adsabs.harvard.edu/abs/2017ARA&A..55...59N} {55, 59}

\bibitem[\protect\citeauthoryear{{Nelson} et~al.,}{{Nelson}
  et~al.}{2019}]{Nelson2019}
{Nelson} D.,  et~al., 2019, \mn@doi [\mnras] {10.1093/mnras/stz2306}, \href
  {https://ui.adsabs.harvard.edu/abs/2019MNRAS.490.3234N} {490, 3234}

\bibitem[\protect\citeauthoryear{Newville et~al.,}{Newville
  et~al.}{2023}]{lmfit}
Newville M.,  et~al., 2023, lmfit/lmfit-py: 1.2.1,
  \mn@doi{10.5281/zenodo.7887568}, \url
  {https://doi.org/10.5281/zenodo.7887568}

\bibitem[\protect\citeauthoryear{{Oppenheimer} \& {Dav{\'e}}}{{Oppenheimer} \&
  {Dav{\'e}}}{2008}]{Oppenheimer2008}
{Oppenheimer} B.~D.,  {Dav{\'e}} R.,  2008, \mn@doi [\mnras]
  {10.1111/j.1365-2966.2008.13280.x}, \href
  {https://ui.adsabs.harvard.edu/abs/2008MNRAS.387..577O} {387, 577}

\bibitem[\protect\citeauthoryear{{Osterbrock} \& {Ferland}}{{Osterbrock} \&
  {Ferland}}{2006}]{Osterbrock2006}
{Osterbrock} D.~E.,  {Ferland} G.~J.,  2006, {Astrophysics of gaseous nebulae
  and active galactic nuclei}.
University Science Books

\bibitem[\protect\citeauthoryear{{{\"O}stlin} et~al.,}{{{\"O}stlin}
  et~al.}{2014}]{Ostlin2014}
{{\"O}stlin} G.,  et~al., 2014, \mn@doi [\apj] {10.1088/0004-637X/797/1/11},
  \href {https://ui.adsabs.harvard.edu/abs/2014ApJ...797...11O} {797, 11}

\bibitem[\protect\citeauthoryear{{P{\'e}roux}, {Nelson}, {van de Voort},
  {Pillepich}, {Marinacci}, {Vogelsberger}  \& {Hernquist}}{{P{\'e}roux}
  et~al.}{2020}]{Peroux2020}
{P{\'e}roux} C.,  {Nelson} D.,  {van de Voort} F.,  {Pillepich} A.,
  {Marinacci} F.,  {Vogelsberger} M.,   {Hernquist} L.,  2020, \mn@doi [\mnras]
  {10.1093/mnras/staa2888}, \href
  {https://ui.adsabs.harvard.edu/abs/2020MNRAS.499.2462P} {499, 2462}

\bibitem[\protect\citeauthoryear{{Pillepich} et~al.,}{{Pillepich}
  et~al.}{2018}]{Pillepich2018}
{Pillepich} A.,  et~al., 2018, \mn@doi [\mnras] {10.1093/mnras/stx2656}, \href
  {https://ui.adsabs.harvard.edu/abs/2018MNRAS.473.4077P} {473, 4077}

\bibitem[\protect\citeauthoryear{{Reichardt Chu} et~al.,}{{Reichardt Chu}
  et~al.}{2022}]{ReichardtChu2022}
{Reichardt Chu} B.,  et~al., 2022, \mn@doi [\mnras] {10.1093/mnras/stac420},
  \href {https://ui.adsabs.harvard.edu/abs/2022MNRAS.511.5782R} {511, 5782}

\bibitem[\protect\citeauthoryear{{Rubin}, {Prochaska}, {Koo}, {Phillips},
  {Martin}  \& {Winstrom}}{{Rubin} et~al.}{2014}]{Rubin2014}
{Rubin} K. H.~R.,  {Prochaska} J.~X.,  {Koo} D.~C.,  {Phillips} A.~C.,
  {Martin} C.~L.,   {Winstrom} L.~O.,  2014, \mn@doi [\apj]
  {10.1088/0004-637X/794/2/156}, \href
  {https://ui.adsabs.harvard.edu/abs/2014ApJ...794..156R} {794, 156}

\bibitem[\protect\citeauthoryear{{Rupke}, {Veilleux}  \& {Sanders}}{{Rupke}
  et~al.}{2005}]{Rupke2005}
{Rupke} D.~S.,  {Veilleux} S.,   {Sanders} D.~B.,  2005, \mn@doi [\apjs]
  {10.1086/432889}, \href
  {https://ui.adsabs.harvard.edu/abs/2005ApJS..160..115R} {160, 115}

\bibitem[\protect\citeauthoryear{{Rupke} et~al.,}{{Rupke}
  et~al.}{2019}]{Rupke2019}
{Rupke} D. S.~N.,  et~al., 2019, \mn@doi [\nat] {10.1038/s41586-019-1686-1},
  \href {https://ui.adsabs.harvard.edu/abs/2019Natur.574..643R} {574, 643}

\bibitem[\protect\citeauthoryear{{Schneider}, {Ostriker}, {Robertson}  \&
  {Thompson}}{{Schneider} et~al.}{2020}]{Schnieder2020}
{Schneider} E.~E.,  {Ostriker} E.~C.,  {Robertson} B.~E.,   {Thompson} T.~A.,
  2020, \mn@doi [\apj] {10.3847/1538-4357/ab8ae8}, \href
  {https://ui.adsabs.harvard.edu/abs/2020ApJ...895...43S} {895, 43}

\bibitem[\protect\citeauthoryear{{Shaban} et~al.,}{{Shaban}
  et~al.}{2022}]{Shaban2022}
{Shaban} A.,  et~al., 2022, \mn@doi [\apj] {10.3847/1538-4357/ac7c65}, \href
  {https://ui.adsabs.harvard.edu/abs/2022ApJ...936...77S} {936, 77}

\bibitem[\protect\citeauthoryear{{Shopbell} \& {Bland-Hawthorn}}{{Shopbell} \&
  {Bland-Hawthorn}}{1998}]{Shopbell1998}
{Shopbell} P.~L.,  {Bland-Hawthorn} J.,  1998, \mn@doi [\apj] {10.1086/305108},
  \href {https://ui.adsabs.harvard.edu/abs/1998ApJ...493..129S} {493, 129}

\bibitem[\protect\citeauthoryear{{Somerville} \& {Dav{\'e}}}{{Somerville} \&
  {Dav{\'e}}}{2015}]{Somerville2015}
{Somerville} R.~S.,  {Dav{\'e}} R.,  2015, \mn@doi [\araa]
  {10.1146/annurev-astro-082812-140951}, \href
  {https://ui.adsabs.harvard.edu/abs/2015ARA&A..53...51S} {53, 51}

\bibitem[\protect\citeauthoryear{{Stanway} \& {Eldridge}}{{Stanway} \&
  {Eldridge}}{2018}]{Stanway2018}
{Stanway} E.~R.,  {Eldridge} J.~J.,  2018, \mn@doi [\mnras]
  {10.1093/mnras/sty1353}, \href
  {https://ui.adsabs.harvard.edu/abs/2018MNRAS.479...75S} {479, 75}

\bibitem[\protect\citeauthoryear{{Steidel}, {Erb}, {Shapley}, {Pettini},
  {Reddy}, {Bogosavljevi{\'c}}, {Rudie}  \& {Rakic}}{{Steidel}
  et~al.}{2010}]{Steidel2010}
{Steidel} C.~C.,  {Erb} D.~K.,  {Shapley} A.~E.,  {Pettini} M.,  {Reddy} N.,
  {Bogosavljevi{\'c}} M.,  {Rudie} G.~C.,   {Rakic} O.,  2010, \mn@doi [\apj]
  {10.1088/0004-637X/717/1/289}, \href
  {https://ui.adsabs.harvard.edu/abs/2010ApJ...717..289S} {717, 289}

\bibitem[\protect\citeauthoryear{{Tumlinson}, {Peeples}  \& {Werk}}{{Tumlinson}
  et~al.}{2017}]{Tumlinson2017}
{Tumlinson} J.,  {Peeples} M.~S.,   {Werk} J.~K.,  2017, \mn@doi [\araa]
  {10.1146/annurev-astro-091916-055240}, \href
  {https://ui.adsabs.harvard.edu/abs/2017ARA&A..55..389T} {55, 389}

\bibitem[\protect\citeauthoryear{{Veilleux} \& {Rupke}}{{Veilleux} \&
  {Rupke}}{2002}]{Veilleux2002}
{Veilleux} S.,  {Rupke} D.~S.,  2002, \mn@doi [\apjl] {10.1086/339226}, \href
  {https://ui.adsabs.harvard.edu/abs/2002ApJ...565L..63V} {565, L63}

\bibitem[\protect\citeauthoryear{{Veilleux}, {Shopbell}, {Rupke},
  {Bland-Hawthorn}  \& {Cecil}}{{Veilleux} et~al.}{2003}]{Veilleux2003}
{Veilleux} S.,  {Shopbell} P.~L.,  {Rupke} D.~S.,  {Bland-Hawthorn} J.,
  {Cecil} G.,  2003, \mn@doi [\aj] {10.1086/379000}, \href
  {https://ui.adsabs.harvard.edu/abs/2003AJ....126.2185V} {126, 2185}

\bibitem[\protect\citeauthoryear{{Veilleux}, {Cecil}  \&
  {Bland-Hawthorn}}{{Veilleux} et~al.}{2005}]{Veilleux2005}
{Veilleux} S.,  {Cecil} G.,   {Bland-Hawthorn} J.,  2005, \mn@doi [\araa]
  {10.1146/annurev.astro.43.072103.150610}, \href
  {https://ui.adsabs.harvard.edu/abs/2005ARA&A..43..769V} {43, 769}

\bibitem[\protect\citeauthoryear{{Veilleux}, {Maiolino}, {Bolatto}  \&
  {Aalto}}{{Veilleux} et~al.}{2020}]{Veilleux2020}
{Veilleux} S.,  {Maiolino} R.,  {Bolatto} A.~D.,   {Aalto} S.,  2020, \mn@doi
  [\aapr] {10.1007/s00159-019-0121-9}, \href
  {https://ui.adsabs.harvard.edu/abs/2020A&ARv..28....2V} {28, 2}

\bibitem[\protect\citeauthoryear{{Westmoquette}, {Smith}, {Gallagher},
  {Trancho}, {Bastian}  \& {Konstantopoulos}}{{Westmoquette}
  et~al.}{2009}]{Westmoquette2009}
{Westmoquette} M.~S.,  {Smith} L.~J.,  {Gallagher} J.~S. I.,  {Trancho} G.,
  {Bastian} N.,   {Konstantopoulos} I.~S.,  2009, \mn@doi [\apj]
  {10.1088/0004-637X/696/1/192}, \href
  {https://ui.adsabs.harvard.edu/abs/2009ApJ...696..192W} {696, 192}

\bibitem[\protect\citeauthoryear{{Xu} et~al.,}{{Xu} et~al.}{2023}]{xu2023}
{Xu} X.,  et~al., 2023, \mn@doi [arXiv e-prints] {10.48550/arXiv.2301.11498},
  \href {https://ui.adsabs.harvard.edu/abs/2023arXiv230111498X} {p.
  arXiv:2301.11498}

\bibitem[\protect\citeauthoryear{{Zabl} et~al.,}{{Zabl}
  et~al.}{2021}]{Zabl2021}
{Zabl} J.,  et~al., 2021, \mn@doi [\mnras] {10.1093/mnras/stab2165}, \href
  {https://ui.adsabs.harvard.edu/abs/2021MNRAS.507.4294Z} {507, 4294}

\makeatother
\end{thebibliography}




\appendix

\section{Uncertain parameters in minor axis mass outflow rate}
\label{sec:uncertainparams}

In this appendix we discuss the impact of the outflow electron density and velocity profiles on the mass outflow rate in the minor axis region. 

\subsection{Radial Electron Density Profile}
\label{subsec:elecdensity}

The [OII]~$\lambda3729/$[OII]~$\lambda3727$ ratio is a sensitive tracer of the election density above $n_{e}\approx20$~cm$^{-3}$ and this [OII] doublet is within the wavelength range of our observations. Within the galaxy disk, the high surface brightness and signal-to-noise ratio of the spectra, as well as the low velocity dispersion of the gas make distinguishing the peaks of the doublet possible. To estimate the electron density in the disk, we sum the spectra within a region with a minor axis height of 1~kpc, and a major axis width equal to the width of the galaxy disk, centered on the galaxy. Within this region we measure an electron density of $n_{e}<32$~cm$^{-3}$. 

At distances above the plane of the disk the complexity of the line shapes makes deblending the [OII] doublet unreliable. Additionally, at even moderate minor-axis offsets the [OII] doublet is not fully resolved in our observations. As stated above, separate peaks can be distinguished within the disk, but within the kinematically-complex, low surface-brightness outflows the [OII] doublet is difficult to reliably fit with a model that corresponds with physical values for $n_{e}$. The classic outflow model \citep{Chevalier1985} derives a radial electron density profile $n_{e}\propto r^{-2}$. Recent work however suggests this profile may be flatter \citep{Lopez2020, Schnieder2020, Fielding2022}. We are therefore required to adopt an assumption for the gradient in electron density moving away from the galaxy mid-plane. In this analysis, we test three models for the electron density profile, given by: 
\begin{equation}
\label{eq:electrondensityprofile}
    n_{e, z}=n_{e, {\rm max}}\left(\frac{z}{h_{n_{e}}}\right)^{\alpha},
\end{equation}
for $\alpha=0, -1, -2$. Here $n_{e, {\rm max}}=32$~cm$^{-3}$, our disk electron density. We adopt a scale height, $h_{n_{e}}=0.8$~kpc, such that for $\alpha=-1, -2$ the density reaches the maximum value at the disk edge. All points for which $z\leq h_{n_{e}}$ are set to $n_{e}=n_{e, max}$. We assume a constant outflow velocity $v_{\rm out} = 300$~km~s$^{-1}$ (as discussed in Appendix~\ref{subsec:appendixvelocity}). Table~\ref{tab:electrondensitymodels} shows the total calculated electron density in the 90\% minor axis outflow regions for different values of $\alpha$ in Eq.~\ref{eq:electrondensityprofile}. The adoption of a flat electron density profile ($\alpha=0$) results in the lowest mass outflow rate while increasing values of $\alpha$ and thus more rapidly decaying electron density profiles result in greater total mass outflow rates. As the minor-axis electron density profile in Mrk~1486 is not well empirically motivated, and the suggestion in recent work is that this profile is flatter than $\alpha=-2$ \citep{Lopez2020, Schnieder2020, Fielding2022}, we adopt the intermediate value of $\alpha=-1$ in the body of the text. Since this profile is unconstrained in our observations, however, our mass outflow rate figures should be interpreted as relative ones, and will change dramatically under different assumptions about $n_{e}$.

\begin{table}
	\centering
	\caption{Total mass outflow rates for the 90\% minor axis outflow regions for different minor axis electron density models.}
	\label{tab:electrondensitymodels}
	\begin{tabular}{ccc} 
		\hline
		Region & $\alpha$ & $\dot{M}_{\rm out}$($M_{\odot}$~yr$^{-1}$)\\
		\hline
            NW & 0 & 0.5\\
            NW & -1 & 1.2\\
            NW & -2 & 2.9\\
            SE & 0 & 0.7\\
            SE & -1 & 1.3\\
            SE & -2 & 3.1\\
		\hline
	\end{tabular}
\end{table}

\subsection{Minor Axis Outflow Velocity Scale Height}\label{subsec:appendixvelocity}

As Mrk~1486 is edge-on, the outflow velocity in the minor axis outflow regions is poorly constrained. Here we experiment with two radial velocity profiles for the minor axis outflow. The first profile has a constant velocity with radial position, $z$, while the second has a velocity, $v(z)$, given by:
\begin{equation}
\label{eq:outflowvelocityapp}
    v_{\rm out}(z) = v_{\rm out, max}\left(1-\exp\left(\frac{-z}{h_{v_{\rm out}}}\right)\right).
\end{equation}
\citet{Chisholm2015} and \citet{Heckman2015} suggest a range of outflow velocities $150-450$~km~s$^{-1}$. In this analysis, we adopt the median value of $300$~km~s$^{-1}$ as the velocity in the constant scenario, and as $v_{\rm out, max}$ in the variable scenario. In the variable outflow velocity profile (Eq.~\ref{eq:outflowvelocityapp}) the choice of scale height, $h_{v_{\rm out}}$, is not well motivated by our observations, and there is little precedent for this in the literature. We thus calculate mass outflow rates at the 90\% outflow width for a range of scale heights. Here we assume an electron density profile consistent with Eq.~\ref{eq:electrondensityprofile} for $\alpha=-1$ and $n_{e, {\rm max}}=32$~cm$^{-3}$ per Appendix~\ref{subsec:elecdensity}. These mass outflow rates are shown in Table~\ref{tab:scaleheightmassout}. The assumption of constant outflow velocity ($h_{v_{\rm out}}$=0) results in an upper bound for the measured mass outflow rate. Increasing values for $h_{v_{\rm out}}$ results in decreases in the calculated mass outflow rate, however this dependence is very weak. The mass outflow rate in the NW outflow region is largely insensitive to reasonable increases in $h_{v_{\rm out}}$, while the SE outflow region requires a large value of $h_{v_{\rm out}}=3$~kpc to have a significant impact on the measured mass outflow rate. As the mass outflow rate is only weakly dependent on this quantity, we adopt the simple, constant velocity model, with $v_{\rm out}=300$~km~s$^{-1}$ in the body of the text. As noted above this makes our mass outflow rate values upper bounds in terms of the selected velocity profile.

\begin{table}
	\centering
	\caption{Total mass outflow rates for the minor axis outflow regions for different values of $h_{v_{\rm out}}$.}
	\label{tab:scaleheightmassout}
	\begin{tabular}{ccc} 
		\hline
		Region & $h_{v_{\rm out}}$ (kpc) & $\dot{M}_{\rm out}$($M_{\odot}$~yr$^{-1}$)\\
		\hline
            NW & 0 & 1.2\\
            NW & 0.5 & 1.2\\
            NW & 1.0 & 1.1\\
            NW & 1.5 & 1.1\\
            NW & 2.0 & 1.1\\
            NW & 2.5 & 1.0\\
            NW & 3.0 & 1.0\\
            SE & 0 & 1.3\\
            SE & 0.5 & 1.3\\
            SE & 1.0 & 1.1\\
            SE & 1.5 & 1.0\\
            SE & 2.0 & 0.8\\
            SE & 2.5 & 0.7\\
            SE & 3.0 & 0.7\\
		\hline
	\end{tabular}
\end{table}

\bsp	
\label{lastpage}
\end{document}